\documentclass[prb, twocolumn,superscriptaddress]{revtex4-1}
\usepackage{amsmath,amssymb,bm}
\usepackage{hyperref}
\usepackage{graphicx}
\usepackage{epstopdf}
\usepackage{latexsym}
\usepackage{subfigure}
\usepackage[usenames, dvipsnames]{color}
\usepackage{natbib}
\usepackage{braket}
\usepackage{float}
\usepackage[normalem]{ulem}
\usepackage{comment}
\usepackage{mathtools}
\usepackage{array}
\usepackage{tabu}
\usepackage{multirow}

\newcommand\redsout{\bgroup\markoverwith{\textcolor{red}{\rule[0.5ex]{2pt}{0.4pt}}}\ULon}
\newcommand\bluesout{\bgroup\markoverwith{\textcolor{blue}{\rule[0.5ex]{2pt}{0.4pt}}}\ULon}

\newcommand{\ADhide}[1]{{}}
\newcommand{\SPhide}[1]{{}}

\begin{document}
\title{$SU(3)$ fermions in a three-band Graphene-like model}
\author{Ankur Das}
\affiliation{Department of Physics and Astronomy, University of Kentucky,
Lexington, KY 40506, USA}
\author{Sumiran Pujari}
\affiliation{Department of Physics, Indian Institute of Technology Bombay, Mumbai, MH 400076, India}
\begin{abstract}
Two-dimensional Graphene is fascinating because of its unique electronic properties.  From a fundamental perspective, one among them is the geometric phase structure near the Dirac points in the Brillouin zone, owing to the $SU(2)$ nature of the Dirac cone wave functions.  We ask if there are geometric phase structures in two dimensions which go beyond that of a Dirac cone. Here we write down a family of three-band continuum models of non-interacting fermions which have more intricate geometric phase structures. This is connected to the $SU(3)$ nature of the wavefunctions near three-fold degeneracies. 
We also give a tight-binding free fermion model on a two-dimensional Graphene-like lattice where the three-fold degeneracies are realized at fine-tuned points. Away from them, we obtain new ``three-band" Dirac cone structures with associated non-standard Landau level quantization, 
whose organization is strongly affected by the non-$SU(2)$ or beyond-Dirac geometric phase structure of the fine-tuned points.
\end{abstract}
\maketitle

\section{Introduction}
\label{sec:intro}
Geometric phases and related concepts have proved invaluable in understanding quantum mechanical phenomena which depend on the analytic structure of a parameterized Hilbert space, ranging from magneto-electric phenomena to topological insulators.\cite{Vanderbilt2018, Di_Chang_Niu2010} Another striking domain is the quantum Hall effect that has been very instructive to modern condensed matter.\cite{Di_Chang_Niu2010} For weakly correlated electrons, they manifest in the Hilbert space of the Bloch wavefunctions of a band solid, where the parameter is the reciprocal crystal momentum. Berry phase is routinely used to such quantify geometric phases.\cite{Berry1984} There are other quantifiers as well, e.g. the famous TKNN invariant is used in integer quantum Hall effect to quantify the geometric phase structure of filled gapped bands.\cite{TKNN1982} 

A very familiar example of a Berry phase effect in two dimensions (2$d$) is that of Graphene. \cite{Vafek_Vishwanath2014}
The Dirac cone spectra associated with mono-layer Graphene has a 
non-trivial geometric phase structure for each cone.
Specifically, there is a Berry phase of $e^{i \pi}$ when one circuits around the cone.
In presence of magnetic field and consequent Landau level formation, 
this structure can manifest sometimes
as new Hall conductance plateaus. \cite{Zhang_etal2006}
Multi-layer Graphene \cite{Novoselov_etal2006}  is another example that hosts 
other multiples of $\pi$ Berry phases around its Dirac-like degeneracies.

For the above examples, 
there is an argument going back to Berry's original article \cite{Berry1984} 
that shows why we get multiples of $\pi$ (rather half-integral multiples of $2\pi$). 
When one considers the 
most general Hamiltonian that can characterize any 
two-fold degeneracy as happens in Dirac-like degeneracies, one obtains a $SU(2)$ 
matrix $H= \sum_{i \in \{1,2,3\}} \lambda_i \hat{\sigma_i}$.
\cite{SU2_simple_example}
The geometric phase 
in this case, is half the solid angle subtended 
by the circuit in the parameter space at the degeneracy point. Since in the preceding 
we have in general three parameters, for a non-fine-tuned degeneracy as a function of 
$2d$ crystal momentum in a $2d$ Bloch Hamiltonian, we need symmetries to 
reduce down to two parameters.\cite{graphene_symmetries}
\SPhide{ADD footnote e.g. Graphene symmetries
... AD: ``e.g. in Graphene, space inversion, and time reversal  
get rid of one of the parameters, REF to Vafek-Vishwanath"}
Once so restricted to two parameters, we can only get  
multiples of $\pi$.
This is the usual Dirac-like geometric phase structure in two dimensions.

The above argument motivates the starting point of this paper.  
Can there be two-dimensional non-interacting electronic band structures that host non-Dirac geometric phase structures? 
Since we have just argued that this possibility is absent for two-fold degeneracies, we have to look beyond them. 
The simplest generalization can then be a three-fold degeneracy. Thus we start by 
writing down a natural three-band generalization of the Dirac cone continuum Hamiltonian 
(Eq. \ref{eq:3bandcontham}). 
We find that this particular generalization 
indeed 
hosts a more intricate geometric phase structure than the two-fold Dirac cone wavefunctions. 
Even at the level of the spectrum, there are two-fold line-degeneracies that emanate
from the three-fold degenerate point in the parameter space.
Because of these line-degeneracies, and thereby a lack of adiabaticity
in the usual sense \cite{Berry1984}, 
computing Berry phase around the three-fold degeneracy 
is formally
problematic.
\cite{line_degeneracy_berry_phase}

This leads us to describe this  
geometric phase structure using a triplet of indices which
tracks how many times each member of the triplet 
(parameter, the Hamiltonian, the eigenfunctions) individually
wind back to themselves as the system winds back to 
itself once, as we keep circuiting around the degeneracy point. 
Here, by the system, we mean the collection of the parameters, the Hamiltonian and all eigenfunctions.
For our purpose, it proves useful to employ 
this method to classify the geometric phase structure near a multi-fold degenerate point in 2$d$
especially in the presence of line-degeneracies.
This quantifier may be thought of as a conceptual generalization of the pseudospin winding number.
\cite{Park_Marzari2011}

Analyzing this toy model further from the point of view of what space symmetries can guarantee this kind of a three-fold degeneracy, we find an interesting $SU(3)$ group structure near the three-fold degeneracy. 
It turns out that the above mentioned beyond-Dirac-like geometric phase structure obtains at fine-tuned points. Away from them, we get novel 
three-band Dirac-cone spectra that are a consequence of being ``adiabatically"
connected to this kind of three-fold degeneracy that we have found. We go on to write a tight-binding model which is hosted on a hexagonal lattice similar to Graphene but with three basis sites per unit cell. 
This still preserves the $SU(3)$ structure near three-fold degenerate points, now with an additional valley index arising similar to Graphene. 
Curiously the two-fold line degeneracies mentioned before connect 
the two valleys on a non-contractible loop in the Brillouin zone.

As a contrast to the toy model introduced above, we consider another simple three-band generalization of the Dirac cone as in Eq. \ref{eq:3bandcontham_1} with a three-fold degeneracy. For this case, one again finds a Dirac-like geometric phase structure. This kind of three-fold degeneracy can be thought of as being in the spin-1 representation of $SU(2)$, which is again why we get Berry phase that are multiples of $\pi$. Thus, the $SU(3)$ structure of our toy model is intimately related to its non-Dirac-like geometric phase structure.
Classifying non-trivial multi-fold degeneracies in electronic
band structure as representations of certain groups (constrained by space symmetries) is a powerful point of view. 
\cite{Lin_Liu2015,Bradlyn_etal2016}
In $2d$, there are several works which have considered three-fold degeneracies.
Green \emph{et al} \cite{Green_Santos_Chamon2010}  considered band structures 
resulting from putting microscopic fluxes in the hexagonal and 
Kagome lattices where they found a three-fold degeneracy. Here the fermions are in the spin-1 
representation of the $SU(2)$ group, but their primary motivation was to find flat band structures. 
Subsequently, many other cases of spin-1 $SU(2)$ three-fold degeneracies have been reported.
\cite{Dora_Kailasvuori_Moessner2011,Lan_etal2011,Urban_etal2011,Wang_etal2013,Raoux_etal2014,
Giovannetti_etal2015,
Palumbo_Meichanetzidis2015, Xu_Duan2017, Wang_Yao2018}
Ref. \onlinecite{Raoux_etal2014}'s $\alpha$-$\mathcal{T}_3$ model
on the Dice lattice
is noteworthy because it accommodates an 
interpolation between spin-$\frac{1}{2}$ and spin-1 Dirac fermions.
We also note that in $3d$ which is not our focus, 
three-fold degeneracies have garnered tremendous interest recently, where
they have been sometimes dubbed as triple point fermions. 
\cite{Lv_etal2017,Wrinkler_etal2016,Weng_etal2016a,
Zhu_etal2016,Weng_etal2016b,Chang_etal2017a,Fulga_Stern2017,
Chang_etal2017b,Zhong_etal2017,Yu_Yan_Liu2017,Zhang_etal2017,Yang_etal2017}
All these cases 
correspond to fermions being in the spin-1 representation of $SU(2)$
in the majority
(however, see Ref. \onlinecite{Chang_etal2017a} for an example where
the geometric phase structure goes beyond Weyl/Dirac in 3$d$).

In our $2d$ toy model, unlike the above,
the fermions are in the fundamental representation of the $SU(3)$ group.  
Going away from this fine-tuned model, we get multiple two-fold Dirac cones for generic directions and a three-fold point-degeneracy for some fine-tuned directions. But the $SU(3)$ nature of the toy model manifests itself in the way the various two-fold and three-fold degeneracies are organized to accommodate the $SU(3)$ geometric phase structure. Thus, the presence of the fine-tuned $SU(3)$ point controls the various Dirac cone structures obtained in its vicinity.

The outline of the paper is as follows:
In Sec. \ref{sec:3band}, we write down
a three-band generalization of the Dirac cone Hamiltonian,
and discuss its beyond-Dirac geometric phase structure.
In Sec. \ref{sec:su3}, we study the construction of such
generalizations using symmetries. This gives a family
of three-band Hamiltonians where the fermions transform
in the fundamental representation of $SU(3)$. In Sec. \ref{subsec:various_cases},
we categorize the band structures for various cases of this family of Hamiltonians.
In Sec. \ref{sec:lattice}, we give a lattice model realization
of the above Hamiltonians, and a brief numerical study of the effect
of uniform magnetic field on this non-interacting system.
We conclude with a summary and outlook in Sec. \ref{sec:conclusion}.

\section{Three Band Construction}
\label{sec:3band}
To set the stage, we recall that the low-energy physics of Graphene
is obtained from a two-band lattice hopping model \cite{Wallace1947, Vafek_Vishwanath2014}
which gives two distinct Dirac cones or valleys in the Brillouin zone of the underlying
triangular Bravais lattice. Our convention is to choose the location of valleys
as $\mathbf{K}=\left(\frac{4\pi}{3},0\right)$ and $\mathbf{K}'=\left(-\frac{4\pi}{3},0\right)$ (unit length is 
set by separation
between two neighboring unit cells, and $\mathbf{K}$, $\mathbf{K}'$ are related by a reflection
across the $y$-axis). 
Near one of these valleys in energy units of $\hbar v_F=1$, one can write down the familiar continuum Hamiltonian
\begin{equation}
    H_K^\text{Dirac}(\mathbf{p})=\left(\begin{matrix} 
            0 & p_x-i p_y \\
            p_x+i p_y & 0 
           \end{matrix}\right)
    \label{eq:graphham}
\end{equation}
where $\mathbf{p}$ is the expansion variable near $\mathbf{K}$ with the full crystal momentum being $\mathbf{K} + \mathbf{p}$.
Its eigensystem is 
\begin{align}
\epsilon_1(\mathbf{p})= +p  \: ; \: & \: \: v_1(\mathbf{p})=\frac{1}{\sqrt{2}} \left( e^{-i \theta_{\mathbf{p}}}, 1 \right)^T  \nonumber \\
\epsilon_2(\mathbf{p})= -p  \: ; \: & \: \:  v_2(\mathbf{p}) =\frac{1}{\sqrt{2}} \left( -e^{-i \theta_{\mathbf{p}}}, 1 \right)^T 
\end{align}
where $p=\sqrt{p^2_x + p^2_y}$, and $\theta_{\mathbf{p}} = \arctan \left( \frac{p_y}{p_x} \right)$.
This gives the familiar Berry phase of $e^{i \pi}$ as the parameter $\mathbf{p}$ (and thereby
the angular variable $\theta_{\mathbf{p}}$) winds around once
about the two-fold degenerate point. We note here that during this winding, the full system -- comprising the 
parameter $\mathbf{p}$, the Hamiltonian $H^{\text{Dirac}}_K(\mathbf{p})$, and all eigenvectors 
$\{v_i(\mathbf{p})\}$ -- winds around once as well.

Now, we come to our primary object of interest in this paper.
Consider the following three-band generalization of the continuum Dirac Hamiltonian
$H^{\text{Dirac}}_K(\mathbf{p})$ recalled above, 
\begin{equation}
    H^{3A}_K(\mathbf{p})=\left(\begin{matrix} 
            0 & p_x-i p_y & p_x-ip_y \\
            p_x+i p_y & 0 & p_x+ip_y\\
            p_x+ip_y & p_x-ip_y & 0
           \end{matrix}\right)
        \label{eq:3bandcontham}
\end{equation}
where the subscript $K$ refers again to a valley index anticipating the lattice model
realization of the above in Sec. \ref{sec:lattice}. The eigensystem of $H^{3A}_K(\mathbf{p})$
is
\begin{subequations}
\begin{align}
\epsilon_1^{3A}(\mathbf{p}) & =-2 p\cos\left(\frac{\theta_{\mathbf{p}}+\pi}{3}\right)
\: ; \nonumber \\
v_1^{3A}(\mathbf{p}) & =\frac{1}{\sqrt{3}}
\left(\omega^2 e^{-i \frac{2\theta_{\mathbf{p}}}{3}} \hspace{5mm}
\omega~e^{i \frac{2\theta_{\mathbf{p}}}{3}} \hspace{5mm} 1
\right)^T
\end{align}
\begin{align}
\epsilon_2^{3A}(\mathbf{p}) & = 2 p \cos\left(\frac{\theta_{\mathbf{p}}}{3}\right)
\: ; \nonumber \\
v_2^{3A}(\mathbf{p}) & = \frac{1}{\sqrt{3}}
                \left( e^{-i \frac{2\theta_{\mathbf{p}}}{3}}  \hspace{5mm}
                e^{i \frac{2\theta_{\mathbf{p}}}{3}} \hspace{5mm} 1
                \right)^T
\end{align}
\begin{align}
\epsilon_3^{3A}(\mathbf{p})  &=-2 p \cos\left(\frac{\theta_{\mathbf{p}}-\pi}{3}\right)
\: ; \nonumber \\
v_3^{3A}(\mathbf{p}) & = \frac{1}{\sqrt{3}}
\left( \omega~e^{-i \frac{2\theta_{\mathbf{p}}}{3}} \hspace{5mm}
                 \omega^2 e^{i \frac{2\theta_{\mathbf{p}}}{3}} \hspace{5mm} 1
                \right)^T
\end{align}
\label{eq:3band_eigensystem}
\end{subequations}
where $\omega = e^{i \frac{2\pi}{3}}$, $\omega^2= e^{-i \frac{2\pi}{3}}$
are the complex cube roots of unity.

\begin{figure}
    \centering
    \includegraphics[width=0.8\linewidth]{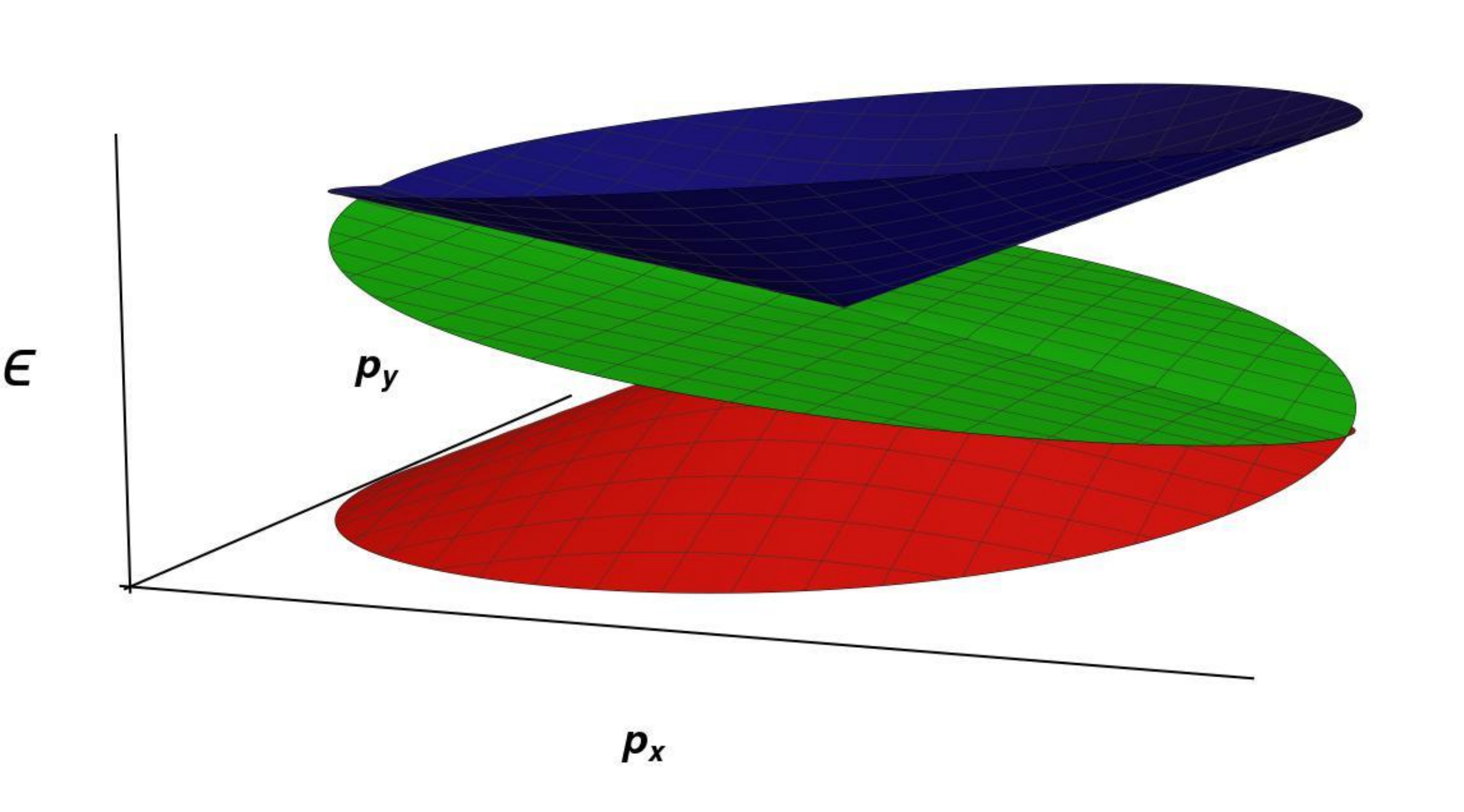}
\caption{The dispersion for $H^{\text{3A}}_K$  (Eqn.\ref{eq:3bandcontham})}
    \label{fig:contband}
\end{figure}

\begin{figure}
    \centering
    \includegraphics[width=0.7\linewidth]{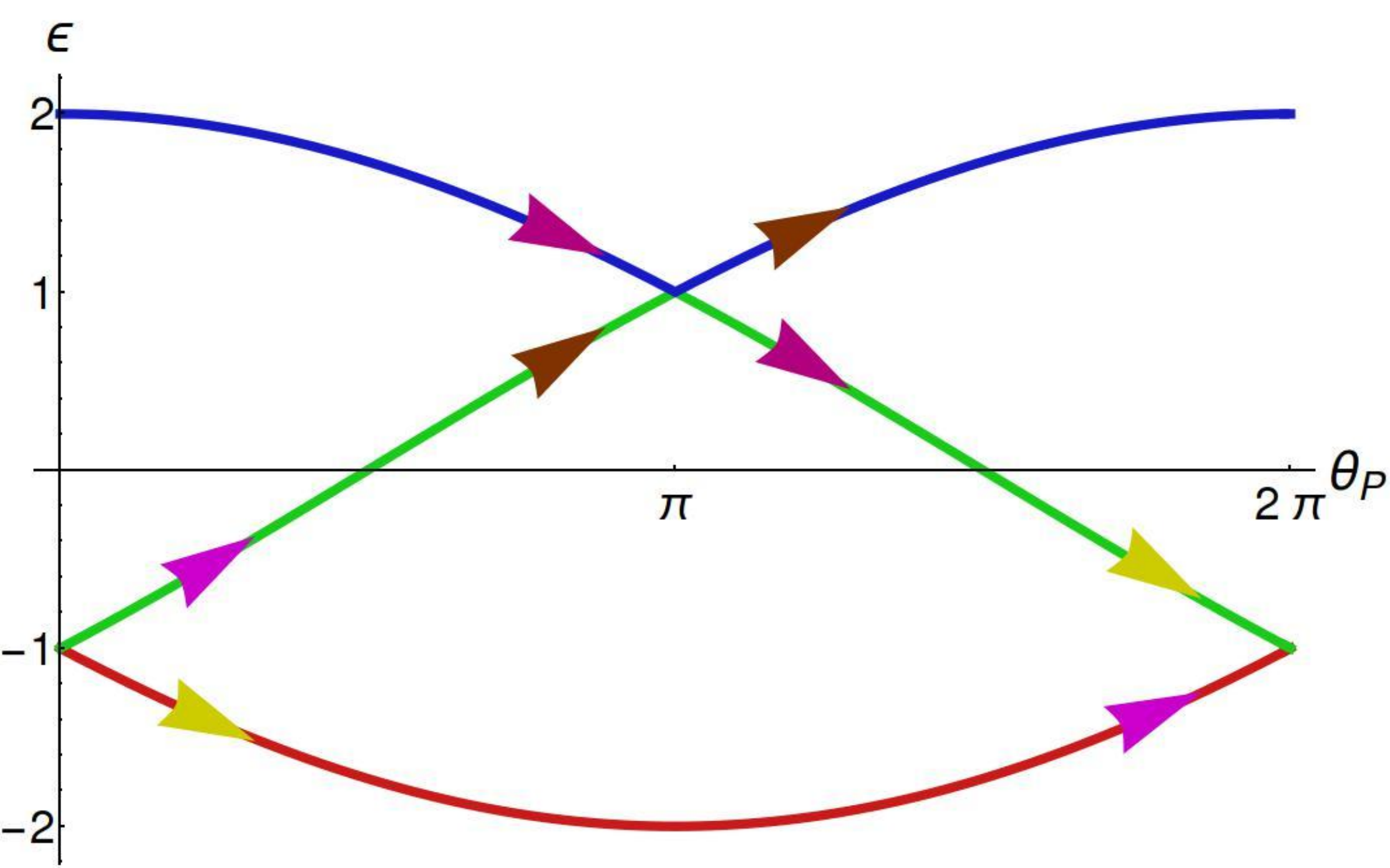}
    \caption{The dispersion for $H^{\text{3A}}_K$  (Eqn.\ref{eq:3bandcontham}) 
at a constant magnitude of momentum ($p=1$) about $\mathbf{p}=(0,0)$ (see Eqn.\ref{eq:3band_eigensystem}). The arrows are a guide to move across the bands in a smooth way, 
where the rule is to follow the same colored arrows while crossing the line degeneracies.}
    \label{fig:contbandang}
\end{figure}

The dispersion near the three-fold degeneracy is shown in Fig.\ref{fig:contband} and
Fig. \ref{fig:contbandang} (for fixed $p=1$). It is
linear in $p$ similar to a Dirac cone, however, now there are a pair of two-fold line
degeneracies that emanate from the three-fold degeneracy outwards in the opposite directions
along the $p_x$-axis. This already gives us a sense of the non-Dirac geometric phase
structure of $H_K^{3A}$.

The source of the non-Dirac geometric phase structure actually lies in the 
$e^{i \theta_{\mathbf{p}}/3 }$ and $e^{i 2 \theta_{\mathbf{p}}/3 }$ terms in $H^{3A}_K$'s eigensystem Eq. \ref{eq:3band_eigensystem}. 
The analytic structure of these terms is qualitatively
different than $e^{i  \theta_{\mathbf{p}} }$ that appears in the 
eigensystem of $H^{\text{Dirac}}_K$.
$z = r e^{i  \theta}$
is an analytic function everywhere on the complex plane,
whereas $z^{1/3} = r^{1/3} e^{i \theta/3}$
has branch-cuts in the complex plane, and needs three Riemann surfaces
to embed the function in an analytic way. The two-fold line degeneracies in Fig. \ref{fig:contband}
are representing these branch-cuts in the following way: we can rewrite the eigenvalues as 
$\bigg\{
\epsilon_1^{3A}(\mathbf{p}) = 2p \: \text{Re}\left[\omega^2 \sqrt[3]{e^{i  \theta_{\mathbf{p}} }}\right],
\epsilon_2^{3A}(\mathbf{p}) = 2p \: \text{Re}\left[ \sqrt[3]{e^{i  \theta_{\mathbf{p}} }}\right],$
$
\epsilon_3^{3A}(\mathbf{p}) = 2p \: \text{Re}\left[ \omega \sqrt[3]{e^{i  \theta_{\mathbf{p}} }}\right]
\bigg\}
$
which are essentially the Riemann surfaces of the complex cube root. We also show this in 
a different way in Fig. \ref{fig:contbandang} through pairs of colored 
arrows which indicate the rule to
move among the bands across the line degeneracies in order to move in an 
analytically smooth way.

\begin{center}
\begin{table}
\begin{tabular}{|m{1cm}|p{1cm}|p{1cm}|p{1cm}|p{2.5cm}|}
\hline
Bands & Model & $\theta_{\mathbf{p}}$ & $H$ & $v_i$ \\
\hline
\hline
\multirow{2}{*}{2} & Dirac & 1 & 1 & 1(\text{t}), 1(\text{b})\\
\cline{2-5}
& QBT & 1 & 2 & 2(\text{t}), 2(\text{b})\\
\hline
\hline
\multirow{2}{*}{3} & $H_K^{3B}$ & 1 & 1 & 1(\text{t}), 2 (\text{m}), 1(\text{b})\\
\cline{2-5}
& $H_K^{3A}$ & 3 & 3 & 2(\text{t}), 2 (\text{m}), 2(\text{b})\\
\hline
\end{tabular}
\caption{This table represents different band structure and their classification using the
triplet of indices introduced in the main text. 
(t) $\rightarrow$ top, (m) $\rightarrow$ middle, (b) $\rightarrow$ bottom.
QBT above refers to Quadratic Band Touching relevant for bilayer Graphene 
\cite{Novoselov_etal2006}, whose continuum Hamiltonian near a valley
may be written as $\left(\begin{matrix} 
0 & (p_x-i p_y)^2 \\
(p_x+i p_y)^2 & 0 
           \end{matrix}\right)$.
  The first three cases are Dirac-like with the first two of them being two-fold cases,
  while the third case is a three-fold case. The very last case is beyond-Dirac-like.
\label{tab:class}
}
\end{table}
\end{center}

To quantify the geometric phase structure of $H_K^{3A}$,
we can not use Berry phase straightforwardly due to the line degeneracies
which present obstructions while doing a circuit in the parameter space
around the three-fold degeneracy. However, from the above discussion, we
see that to move in an analytically smooth way, we have moved across bands
as shown in Fig. \ref{fig:contbandang} as we make circuits around the three-fold degeneracy. This figure also shows that to analytically
return back to the starting point in energy (modulo $p$) at a non-degenerate point, 
we have to circuit three times around the degeneracy.
During these three circuits, one may easily check that the Hamiltonian (Eq. \ref{eq:3bandcontham}) returns back
to itself thrice as well, while the wavefunctions return back to
themselves only \emph{twice} via Eq. \ref{eq:3band_eigensystem}.

This motivates the following triplet of indices to characterize
 the geometric phase structure around any two-dimensional 
multi-fold degeneracy. We track how many times the parameter ($\theta_{\mathbf{p}}$),
the given Hamiltonian, and all eigenfunctions $v^i(\theta_{\mathbf{p}})$ return to
themselves, as we perform a single circuit around the degeneracy for the full
system comprising the parameter,
the Hamiltonian, and all eigenvectors. This triplet of indices
tracking the individual windings is a quantifier of the geometric phase structure
near the degeneracy. This is summarized in Table \ref{tab:class} for several
different two-fold and three-fold degenerate systems. We see from this table
how our three-band model has a non-trivially different geometric phase structure
than the other cases which are all Dirac-like. Thus our model is an example
of a beyond-Dirac geometric phase structure in two dimensions.
It can also be checked that $\left(\begin{matrix} 
            0 & p_x-i p_y & p_x-ip_y \\
            p_x+i p_y & 0 & p_x-ip_y\\
            p_x+ip_y & p_x+ip_y & 0
           \end{matrix}\right)$
  has the same kind of beyond-Dirac geometric phase structure
  as $H_K^{3A}$.

Our three-band continuum Hamiltonian can be contrasted with a Dirac-like three-band case with 
a three-fold degeneracy of the following form
\begin{equation}
    H^{3B}_K(\mathbf{p})=\left(\begin{matrix} 
            0 & p_x+i p_y & 0 \\
            p_x-i p_y & 0 & p_x+ip_y\\
            0 & p_x-ip_y & 0
           \end{matrix}\right)
        \label{eq:3bandcontham_1}
\end{equation}
Its eigensystem is

\begin{subequations}
        \begin{align}
\epsilon_1^{3B}(\mathbf{p}) & =-\sqrt{2} p
         \: ; \nonumber \\
        v_1^{3B}(\mathbf{p}) & =
\left(
 \frac{1}{2} e^{-2 i \theta_{\mathbf{p}} } \hspace{5mm}
 -\frac{e^{-i \theta_{\mathbf{p}} }}{\sqrt{2}} \hspace{5mm} 
 \frac{1}{2}
                \right)^T
\end{align}
\begin{align}
                \epsilon_2^{3B}(\mathbf{p})  &= 0
\: ; \nonumber \\
v_2^{3B}(\mathbf{p}) & = 
\left( -\frac{e^{-2 i \theta_{\mathbf{p}} }}{\sqrt{2}} \hspace{5mm}
 0 \hspace{5mm}
 \frac{1}{\sqrt{2}} 
\right)^T 
\end{align}
\begin{align}
\epsilon_3^{3B}(\mathbf{p}) & = \sqrt{2} p 
                \: ; \nonumber \\
                v_3^{3B}(\mathbf{p}) & = 
\left( \frac{1}{2} e^{-2 i \theta_{\mathbf{p}} } \hspace{5mm}
 \frac{e^{-i \theta_{\mathbf{p}} }}{\sqrt{2}} \hspace{5mm}
\frac{1}{2} \right)^T
        \end{align}
\label{eq:3band_eigensystem_1}
\end{subequations}

As is evident from the eigensystem above, this case has no branch cut structure 
and no line degeneracies.
It can, in fact, be shown that the geometric phase structure, in this case, is still Dirac-like 
similar to $H_K^{\text{Dirac}}$ but in a three-fold situation.
\cite{SU2_spin1}
We also note how it contrasts with beyond-Dirac-like $H_K^{3A}$ in Table \ref{tab:class}.
\SPhide{Add footnote about wavefunction phase?}
In all of the above, we have fixed our gauge by choosing
the last entry of the wavefunctions' column vectors to be purely real.
The classification in 
Table \ref{tab:class} is independent of this gauge choice, since 
wavefunctions differing by pure phases are physically
equivalent and have the same windings around degeneracies.

\section{$SU(3)$ group structure}
\label{sec:su3}

In this section, we expose the underlying $SU(3)$ group structure in Eq. \ref{eq:3bandcontham} as remarked
in  Sec. \ref{sec:intro}. To set the stage, we remind ourselves that 
for monolayer Graphene with two-fold degeneracies,
a continuum Hamiltonian can be written using the Pauli matrices ($SU(2)$ generators in fundamental representation)
near the degeneracy points in the Brillouin zone. 
It is obtained by expanding around the $\mathbf{K}$ and $\mathbf{K'}$ points
in the Brillouin zone of Graphene and looks like
\begin{equation}
\mathcal{H}^{\text{Dirac}}=\sum_{\mathbf{p}} \hat{c}^\dagger_{\mu \alpha}(\mathbf{p}) \:
H^{\text{Dirac}}_{\mu\alpha,\mu' \alpha'} \: \hat{c}_{\mu \alpha}(\mathbf{p})
\end{equation}
where $\hat{c}^\dagger_{\mu \alpha},\hat{c}_{\mu \alpha}$ are the fermion creation and annihilation
operators for the so-called valley index $\mu \in {\mathbf{K},\mathbf{K'}}$ and sublattice index $\alpha
\in {a,b}$, and 
\begin{equation}
H^{\text{Dirac}}_{\mu\alpha,\mu' \alpha'} =
p_x\left( \tau^3_{\mu\mu'} \otimes \sigma^1_{\alpha\alpha'} \right) +
 p_y \left( \tau^0_{\mu\mu'} \otimes \sigma^2_{\alpha\alpha'} \right)
\label{eq:dirac_rewrite}
\end{equation}
and $\tau^i$ are $SU(2)$ Pauli matrices indexing the two valleys, $\sigma^i$
are $SU(2)$ Pauli matrices 
indexing the two sub-lattices of the Graphene lattice, 
$\mathbf{p}$ is the expansion variable (i.e. the
full crystal momentum is $\mathbf{k} = \mathbf{K} + \mathbf{p}$, etc.), and all the indices
have been shown explicitly. We can rewrite Eq. \ref{eq:dirac_rewrite} concisely by
dropping the explicit indices as $H^{\text{Dirac}}=
        p_x\left( \tau^3 \otimes \sigma^1 \right) +
 p_y \left( \tau^0 \otimes \sigma^2 \right)$.
In this way of saying, our three-band 
continuum Hamiltonian written down in the previous section in Eq. \ref{eq:3bandcontham} 
actually requires \emph{all} the off-diagonal generators of the $SU(3)$ group, i.e.
\begin{align}
H^{\text{3A}}_K(\mathbf{p})
= p_x (\Lambda_1 + \Lambda_4 + \Lambda_6) + p_y ( \Lambda_2 + \Lambda_5 - \Lambda_7 )  
\label{eq:3bandcontham_rewrite}
\end{align}
where we use the Gell-Mann matrices as the $SU(3)$ generators. \cite{Halzen_Martin1984}
On the other hand, 
spin-1 generators of the $SU(2)$ group -- which are a subset of the $SU(3)$ group generators --
suffice for the Hamiltonian in Eq. \ref{eq:3bandcontham_1} \SPhide{To AD: Check if prev sentence correct},
\begin{align}
\mathcal{H}^K_{\text{3B}}(\mathbf{p})= p_x (\Lambda_1 + \Lambda_6) + p_y ( \Lambda_2 + \Lambda_7 )
\label{eq:3bandcontham_1_rewrite}
\end{align}

\vspace{3mm}
This leads us to  ask the following question: Along with time-reversal symmetry, what are the spatial point group symmetries 
that we want to preserve while constructing a general continuum  Hamiltonian in two dimensions which has the above 
$SU(3)$ group structure?
In the case of Graphene, 
the spatial point group symmetries of $\mathcal{C}_3$ ($2\pi/3$ rotation about the center of a hexagonal plaquette),
$\mathcal{C}_2$ (inversion, or equivalently a $\pi$ rotation about the center of a hexagonal plaquette)
and $\mathcal{P}_x$/$\mathcal{P}_y$ (reflections about axes passing through the center of a hexagonal plaquette) 
are sufficient to constrain us in writing down Eq. \ref{eq:dirac_rewrite} as the general continuum Hamiltonian that preserves
these symmetries. \SPhide{To AD: check prev sentence}

\SPhide{to AD, IMP, a footnote about conventions of band indexing and appropriate citations}

For our model, we will start by considering a $\mathcal{C}_2$ symmetry. 
The generic operation of the $\mathcal{C}_2$ can be taken to be
as follows: one sublattice remain unchanged, while the other two sublattices get interchanged (e.g. $a \rightarrow b$,
$b \rightarrow a$, $c \rightarrow c$). 
This operation thus looks like 
\begin{equation}
\mathcal{C}_2 \hat{c}_{\mu \alpha}(\mathbf{p}) \mathcal{C}_2^{-1} = 
\left[ \tau^1_{\mu\mu'} \otimes 
C_{2_{\alpha\alpha'}} \right]
\hat{c}_{\mu'\alpha'}(-\mathbf{p})
\label{eq:c2_symm}
\end{equation}
where
\begin{align}
C_2 = & \left(\Lambda^1-\frac{\Lambda^8}{\sqrt{3}}+\frac{\Lambda^0}{3}\right) 
=  \left(\begin{matrix} 
            0 & 1 & 0 \\
            1 & 0 & 0\\
            0 & 0 & 1
           \end{matrix}\right) 
\end{align}
and we stick with $a \rightarrow b$,
$b \rightarrow a$, $c \rightarrow c$ convention as in example above.
However this choice is not special, and all other conventions -- that interchange two sublattices and keep one sublattice
unchanged -- will give us the same spectrum.
$\mathbf{p} \rightarrow -\mathbf{p}$ in the above because the full crystal momentum changes sign,
i.e. $\mathbf{k} = \mathbf{K} + \mathbf{p}
\rightarrow -\mathbf{k} = -\mathbf{K} - \mathbf{p} = \mathbf{K}' - \mathbf{p}$,
which is essentially the $\tau^1$ operation in the valley index, 
and $\mathbf{p} \rightarrow -\mathbf{p}$
for the fermion operators.

We can easily check the following identities for the $C_2$ matrix ($=C^{-1}_2$) that implements $\mathcal{C}_2$,
\begin{subequations}
\begin{align}
\label{eq:transC2}
C_2 \Lambda^1  C_2 
& = \Lambda^1   \\
C_2 \Lambda^2  C_2 
& = - \Lambda^2   \\
C_2 \Lambda^3 C_2 & = -\Lambda^3 \\
C_2 (\Lambda^4 \pm \Lambda^6) C_2 
& = \pm (\Lambda^4 \pm \Lambda^6)  \\
C_2 (\Lambda^5 \pm \Lambda^7) C_2 
& = \pm (\Lambda^5 \pm \Lambda^7)  \\
C_2 \Lambda^8  C_2 
& = \Lambda^8   
\end{align}
\end{subequations}
At the outset, the list of possible terms are of the form $f_{ij}(\mathbf{p}) \: \tau^i \otimes \Lambda^j$
where $f_{ij}(\mathbf{p})$ is some real function of $\mathbf{p}$.
For local Hamiltonians, we can remove terms in this list which contain
$\tau^1$ or $\tau^2$.
So the generic local Hamiltonian that is invariant under $\mathcal{C}_2$ must be a combination from
a reduced list of terms, e.g. terms that contain $\Lambda^1$ will be of the form
\begin{align}
f^+(\mathbf{p}) \: \tau^0 \otimes \Lambda^1 \: \: \: \: \text{and} \: \:\: \: 
f^-(\mathbf{p}) \: \tau^3 \otimes \Lambda^1
\end{align}
where the superscripts are used designate whether the function is even or odd,
i.e. $f^\pm_{ij}(-\mathbf{p}) = \pm f^\pm_{ij}(\mathbf{p})$. We quickly discuss
the reason that governs the even/odd property of the above functional coefficients $f^-(\mathbf{p})$
and $f^+(\mathbf{p})$. This is a standard argument that is used for Graphene as well. We will do
it for the case of $\tau^0 \otimes \Lambda^1$ as an example.

Under $\mathcal{C}_2$ we have $\tau^0\rightarrow \tau^0$. Thus using Eq.\ref{eq:transC2}, $\tau^0\otimes \Lambda^1\rightarrow \tau^0\otimes \Lambda^1$.
 Then (from here on, we suppress indices unless needed)
\begin{align}
\mathcal{C}_2\mathcal{H}^{3A}\mathcal{C}_2^{-1}&=\sum_\mathbf{p} \hat{c}^\dagger(-\mathbf{p}) f^+(\mathbf{p})(\tau^0\otimes\Lambda^1) \hat{c}(-\mathbf{p}) \nonumber \\
&(\text{changing dummy indices as }\mathbf{p}'=-\mathbf{p}) \nonumber \\
&=\sum_{\mathbf{p}'} \hat{c}^\dagger(\mathbf{p}') f^+(-\mathbf{p}')(\tau^0\otimes\Lambda^1) \hat{c}(\mathbf{p}')
\end{align}
Thus $\mathcal{H}^{3A}$ remains invariant if $f^+(-\mathbf{p})=f^+(\mathbf{p})$. All the other possible terms can be analyzed in a similar way. The full list of terms that are finally
allowed by $\mathcal{C}_2$ following the above considerations are
\begin{align}
f^+(\mathbf{p}) \: \tau^0 \otimes \Lambda^1 \: \: &, \: \: f^-(\mathbf{p}) \: \tau^3 \otimes \Lambda^1 \nonumber \\
g^-(\mathbf{p}) \: \tau^0 \otimes \Lambda^2 \: \: &, \: \: g^+(\mathbf{p}) \: \tau^3 \otimes \Lambda^2 \nonumber \\ 
h^-(\mathbf{p}) \: \tau^0 \otimes \Lambda^3 \: \: &, \: \: h^+(\mathbf{p}) \: \tau^3 \otimes \Lambda^3\nonumber \\
l^+_1(\mathbf{p}) \: \tau^0 \otimes (\Lambda^4 + \Lambda^6) \: \: &, \: \: l^-_1(\mathbf{p}) \: \tau^3 \otimes (\Lambda^4 + \Lambda^6) \nonumber \\
l^-_2(\mathbf{p}) \: \tau^0 \otimes (\Lambda^4 - \Lambda^6) \: \: &, \: \: l^+_2(\mathbf{p}) \: \tau^3 \otimes (\Lambda^4 - \Lambda^6) \nonumber \\
m^+_1(\mathbf{p}) \: \tau^0 \otimes (\Lambda^5 + \Lambda^7) \: \: &, \: \: m^-_1(\mathbf{p}) \: \tau^3 \otimes (\Lambda^5 + \Lambda^7) \nonumber \\
m^-_2(\mathbf{p}) \: \tau^0 \otimes (\Lambda^5 - \Lambda^7) \: \: &, \: \: m^+_2(\mathbf{p}) \: \tau^3 \otimes (\Lambda^5 - \Lambda^7) \nonumber \\
n^+(\mathbf{p}) \: \tau^0 \otimes \Lambda^8 \: \: &, \: \: n^-(\mathbf{p}) \: \tau^3 \otimes \Lambda^8
\label{eq:C2_list}
\end{align}
All odd functions in $\mathbf{p}$ at leading order will be linear 
in $p_x,p_y$, and all even functions in $\mathbf{p}$ at leading order will be constants. 
We are mainly interested in these leading order behaviors.
We will comment on higher order terms when needed.

Next, we consider time reversal symmetry $\mathcal{T}$.
Time reversal operation looks like
\begin{equation}
\mathcal{T} \hat{c}
(\mathbf{p}) \mathcal{T}^{-1} = 
\left[ \tau^1 \otimes 
\Lambda^0 \right]\hat{c}(-\mathbf{p})
\label{eq:t_symm}
\end{equation}
(and $\mathcal{T} \: i \: \mathcal{T}^{-1} = -i$). The list of terms
allowed by time reversal symmetry (following
the same steps as $\mathcal{C}_2$) are
\begin{align}
f^+(\mathbf{p}) \: \tau^0 \otimes \Lambda^1 \: \: &, \: \: f^-(\mathbf{p}) \: \tau^3 \otimes \Lambda^1 \nonumber \\
g^-(\mathbf{p}) \: \tau^0 \otimes \Lambda^2 \: \: &, \: \: g^+(\mathbf{p}) \: \tau^3 \otimes \Lambda^2 \nonumber \\ 
l^+_1(\mathbf{p}) \: \tau^0 \otimes (\Lambda^4 + \Lambda^6) \: \: &, \: \: l^-_1(\mathbf{p}) \: \tau^3 \otimes (\Lambda^4 + \Lambda^6) \nonumber \\
m^-_2(\mathbf{p}) \: \tau^0 \otimes (\Lambda^5 - \Lambda^7) \: \: &, \: \: m^+_2(\mathbf{p}) \: \tau^3 \otimes (\Lambda^5 - \Lambda^7) \nonumber \\
n^+(\mathbf{p}) \: \tau^0 \otimes \Lambda^8 \: \: &, \: \: n^-(\mathbf{p}) \: \tau^3 \otimes \Lambda^8
\label{eq:t_list}
\end{align}

Finally, we consider reflection symmetries, $\mathcal{P}_x$ about $y$-axis and $\mathcal{P}_y$ about $x$-axis.
Their combined operations implements $\mathcal{C}_2$ in two dimensions, i.e. $\mathcal{P}_x \mathcal{P}_y = \mathcal{C}_2$.
Now we know from Eq. \ref{eq:c2_symm} that $\mathcal{C}_2$ implements both $\tau^1$ (valley exchange) 
\emph{and} $C_2$ ($a \leftrightarrow b$, $c \leftrightarrow c$). So the non-trivially different symmetry
operations that $\mathcal{P}_x$/$\mathcal{P}_y$ can do are that one of them implements valley exchange, 
and the other implements $C_2$. Also, we note that under $\mathcal{P}_x$, $\mathbf{K} \leftrightarrow \mathbf{K}'$
and under $\mathcal{P}_y$, $\mathbf{K}$/$\mathbf{K}'$ remain unchanged for our choice of the valley locations
in the Brillouin zone.
Therefore, $\mathcal{P}_x$ implements
valley exchange with the sublattices unchanged, and $\mathcal{P}_y$ implements $C_2$ with the valleys unchanged.
Thus, $\mathcal{P}_x$/$\mathcal{P}_y$ operations look like
\begin{subequations}
\begin{align}
\mathcal{P}_x \hat{c}(\mathbf{p}) \mathcal{P}^{-1}_x = 
\left[ \tau^1 \otimes 
\Lambda^0 \right]\hat{c}(\mathcal{P}_x (\mathbf{p})) 
\label{eq:px_symm} \\
\mathcal{P}_y \hat{c}(\mathbf{p}) \mathcal{P}^{-1}_y = 
\left[ \tau^0 \otimes 
C_{2} \right]
\hat{c}(\mathcal{P}_y (\mathbf{p}))
\label{eq:py_symm}
\end{align}
\end{subequations}
Also, because $\mathcal{P}_x (\mathbf{k}) = (-k_x,k_y)$ and $\mathbf{K} \leftrightarrow \mathbf{K}'$
under $\mathcal{P}_x$, therefore $\mathcal{P}_x (\mathbf{p}) = (-p_x,p_y)$ as well.
Similarly, $\mathcal{P}_y (\mathbf{p}) = (p_x,-p_y)$.

Now we explicitly redo the similar analysis as for $\mathcal{C}_2$ for one term $g^+(\mathbf{p})\tau^3\otimes \Lambda^2$ as an example.
\begin{align}
\mathcal{P}_x\mathcal{H}^{3A}\mathcal{P}_x^{-1}&=\sum_\mathbf{p} -\hat{c}^\dagger(\mathcal{P}_x(\mathbf{p})) g^+(\mathbf{p})(\tau^3\otimes\Lambda^2) \hat{c}(\mathcal{P}_x(\mathbf{p})) \nonumber\\
&(\text{changing dummy indices as }\mathbf{p}'=\mathcal{P}_x(\mathbf{p})) \nonumber \\
&=\sum_{\mathbf{p}'} -\hat{c}^\dagger(\mathbf{p}') g^+(\mathcal{P}_x(\mathbf{p}'))(\tau^3\otimes\Lambda^2) \hat{c}(\mathbf{p})
\end{align}
Thus at the leading order $g^+(\mathbf{p})$ will be zero, but higher order terms such $p_xp_y$ 
will satisfy the relation $g^+(\mathcal{P}_x(\mathbf{p}))=-g^+(\mathbf{p})$ and 
are thus allowed. Similarly $m^+_2(\mathbf{p})$ is zero at leading order.  
Therefore, all the terms in Eq. \ref{eq:t_list}
in principle are allowed by reflections, but restricting 
up to leading order, the Hamiltonian looks like 
\begin{align}
H =  & \:  p_x \: \tau^3\otimes (f^- \Lambda^1+l_1^- (\Lambda^4+\Lambda^6)+n^- \Lambda^8) \:\: + \nonumber \\
& \: p_y \: \tau^0 \otimes (g^- \Lambda^2+ m_2^- (\Lambda^5-\Lambda^7)) \:\: + \nonumber \\
& \: \tau_0\otimes (f^+ \Lambda^1+n^+ \Lambda^8+ l_1^+ (\Lambda^4+\Lambda^6))
\label{eq:final_3bandcontham}
\end{align}
In the above, all the function symbols are now replaced by constants. 
This Hamiltonian for $n^-=n^+=f^+=l_1^+=0$ and $f^-=l^-_1=g^-=m_2^-\neq 0$ is
Eq. \ref{eq:3bandcontham}. As discussed in Sec. \ref{sec:3band}, 
Eq. \ref{eq:3bandcontham} has a three-fold degeneracy 
and two two-fold line degeneracies emanating from it (See Fig.\ref{fig:contband}).

\subsection{Various Band Structure}
\label{subsec:various_cases}

In this subsection, we categorize the finer details of various band structures that result from Eq. \ref{eq:final_3bandcontham}.
For the momentum independent terms in Eq. \ref{eq:final_3bandcontham}, 1)
since $\Lambda^8$ is diagonal, 
this must come from a staggered potential contribution. 
If we assume that all orbitals on $a,b,c$ sites are the same, then (like Graphene) we can 
set this term to zero 
($n^+=0$). The first four cases below correspond to this choice. The final
case considers what happens when $n^+ \neq 0$.
2) The $\Lambda_1,\Lambda_4,\Lambda_6$ 
are off-diagonal. Thus, $l^+_1 \neq f^+$ implies a difference
in the $ab$ and $ac,bc$ hoppings (within an unit cell). 
We can measure this deformation in hopping strengths with respect to $ab$ hopping strength, 
i.e. $f^+$ may be set to $0$ without loss of generality.

\begin{figure}
\centering
\includegraphics[width=0.7\linewidth]{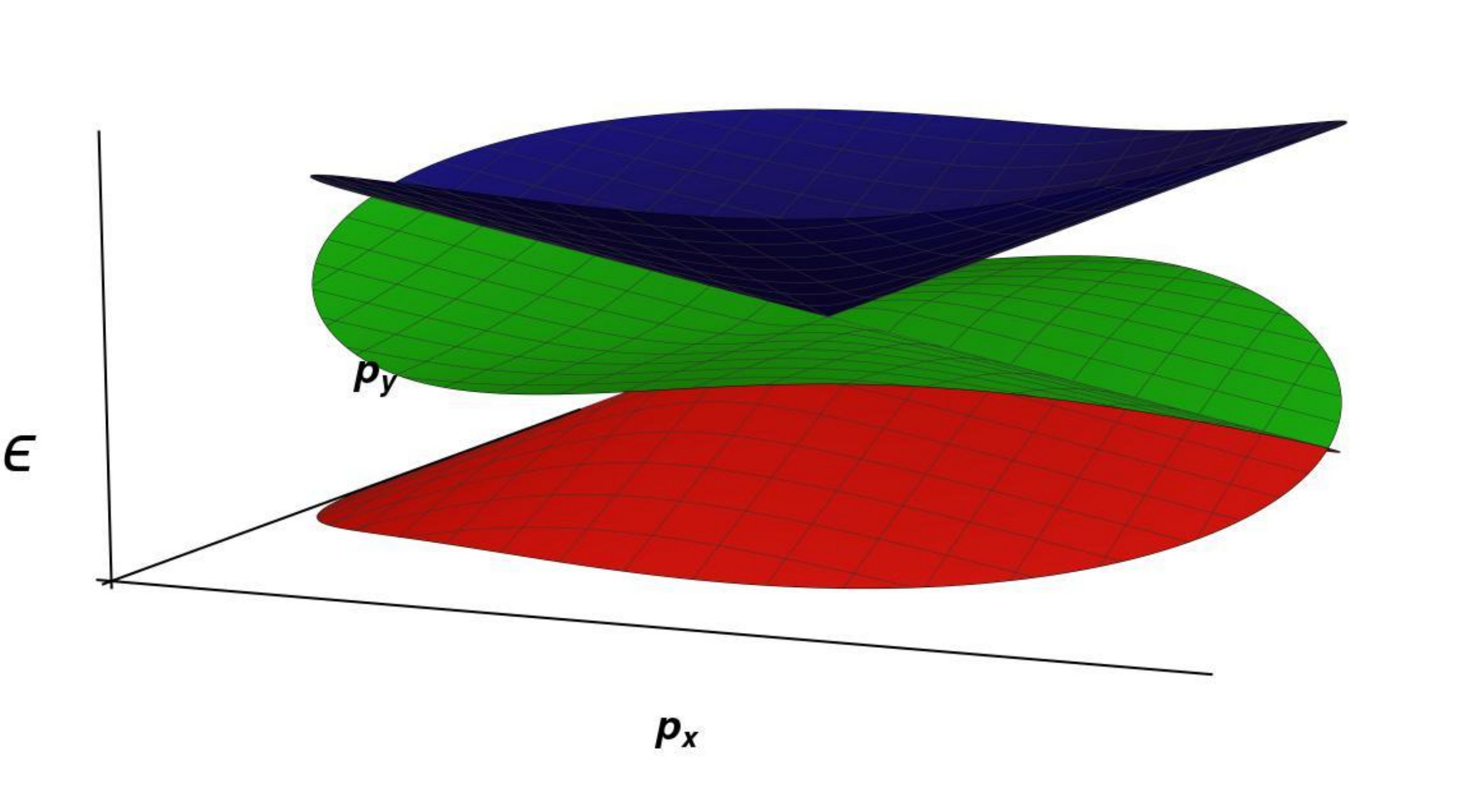}
\caption{Case 1: $l_1^+=0,n^-=0,f^-=l_1^-$
In the above, $g^- \neq f^-$ and $m^-_2 \neq l^-_1$. When all of these quantities are 
equal to each other, we obtain the special case as in Fig. \ref{fig:contband} corresponding
to our starting point $H^{\text{3A}}_K$.}
\label{fig:3Bandgenlindig}
\end{figure}
\begin{figure}
\centering
\includegraphics[width=0.7\linewidth]{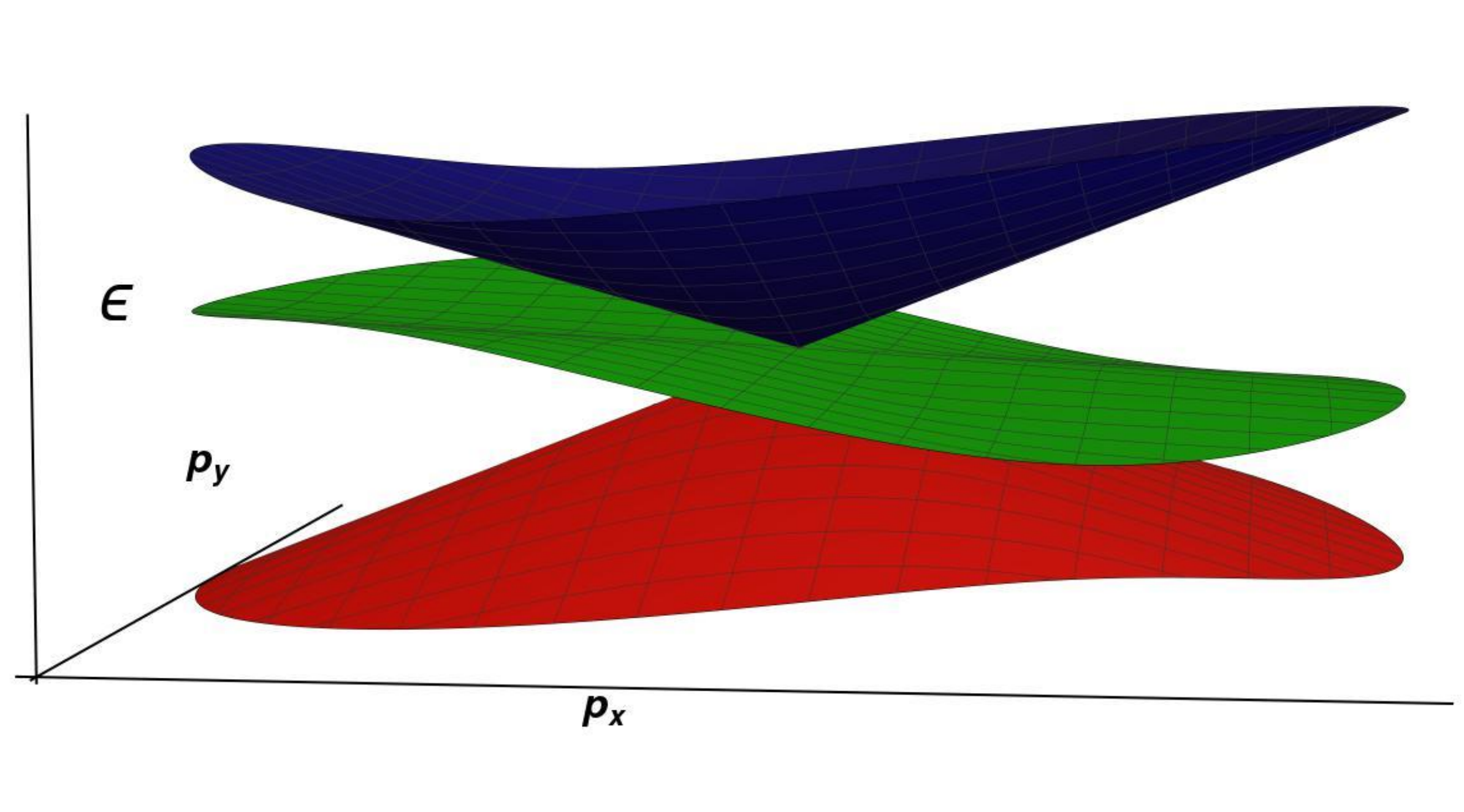}
\caption{Case 2: Representative picture of the effect of $f^-\neq l_1^-$ and/or $n^-\neq 0$ for $l_1^+=0$.}
\label{fig:l8}
\end{figure}
\begin{figure}
\centering
\includegraphics[width=0.7\linewidth]{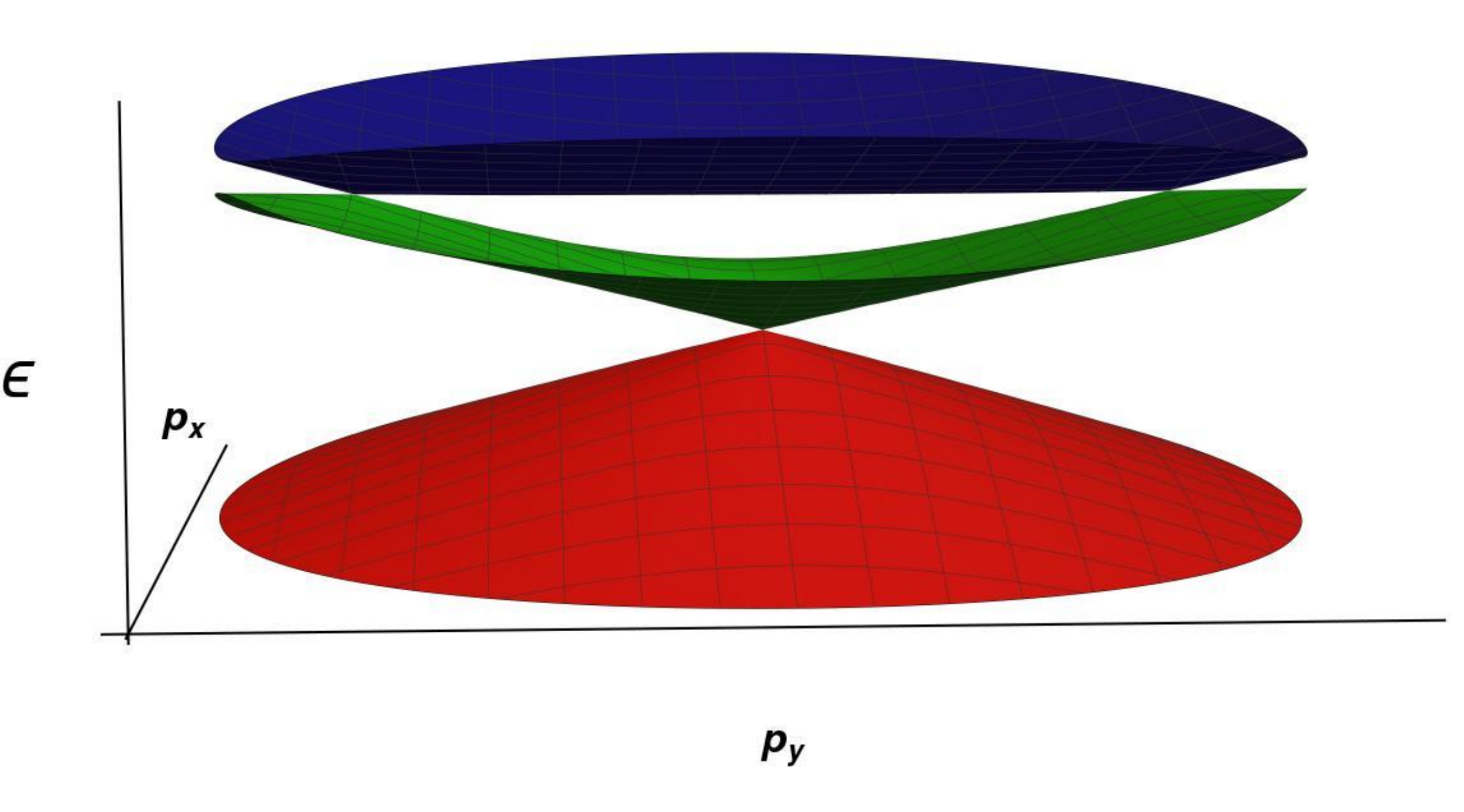}
\caption{Case 3a: $l^+_1 \neq 0$ 
$n^- = 0$, $f^- = l_1^-$.}
\label{fig:3Bandtouchgen}
\end{figure}

Case 1\label{case:1} ($l^+_1=0, n^-=0, f^- = l^-_1$):
This is the base case
with two-fold line degeneracies along $p_y=0$, and three-fold degenerate point at $p_x=p_y=0$.
See Fig.\ref{fig:3Bandgenlindig}.

Case 2\label{case:2} ($l^+_1=0$ AND ($n^- \neq0$ OR $f^- \neq l^-_1$)):
The line degeneracies now  go away, and we end up with only one three band degenerate point (see Fig.\ref{fig:l8}). 
(We note here that this corresponds to the triplet of indices 1 ($\theta_{\mathbf{p}}$), 1 ($H$) and 2(t), 0(m), 2(b).)

Case 3a\label{case:3a} ($l^+_1 \neq 0$, $n^- = 0$, $f^- = l^-_1$):
This is shown in Fig.\ref{fig:3Bandtouchgen}.
Here, the top and middle bands touch each other linearly at two points, while
the middle and bottom bands touch each other linearly at one point.
For the top two bands, the line connecting the degeneracy point is completely flat. 

\begin{figure}
\includegraphics[width=0.7\linewidth]{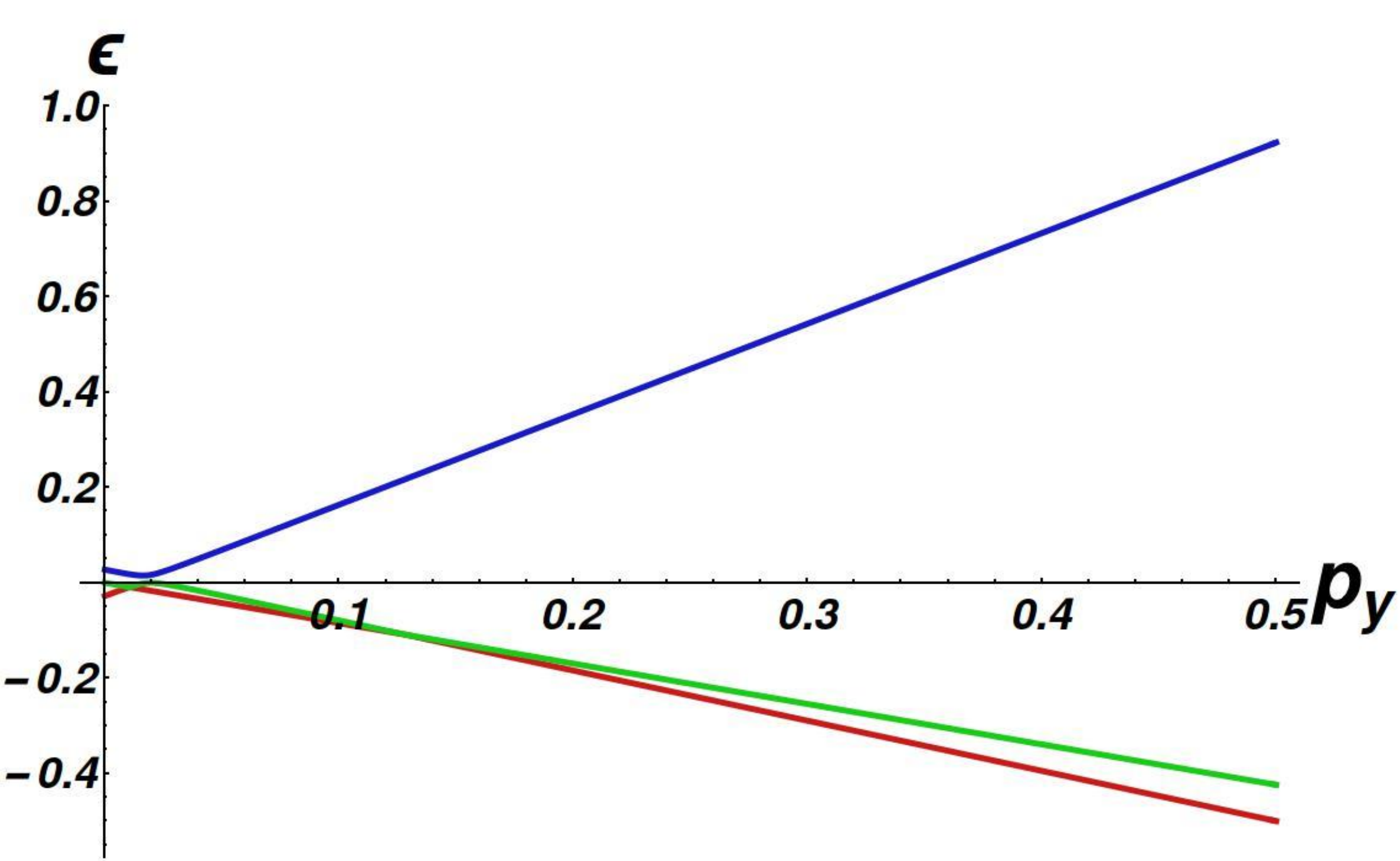}
\includegraphics[width=0.7\linewidth]{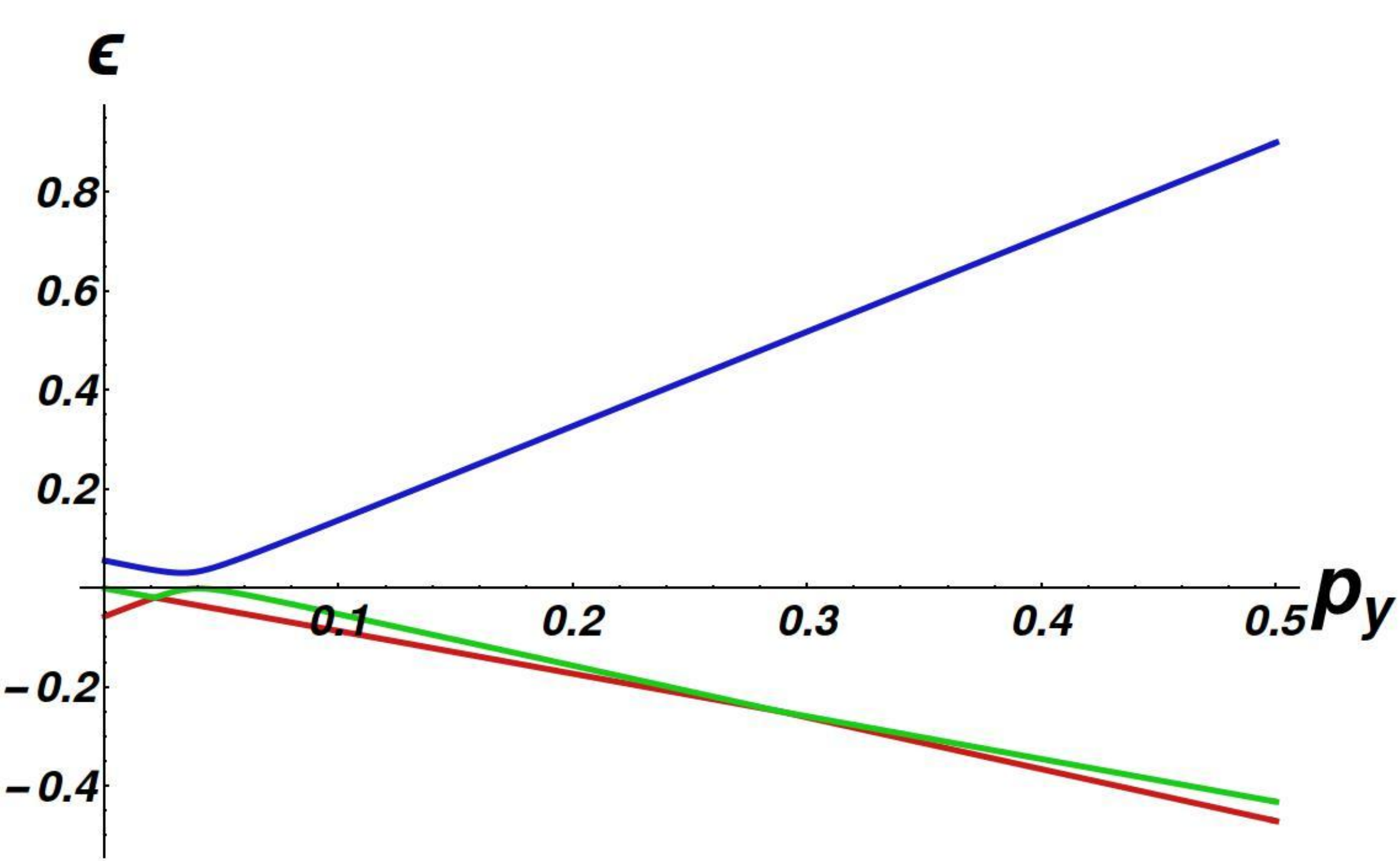}
\includegraphics[width=0.7\linewidth]{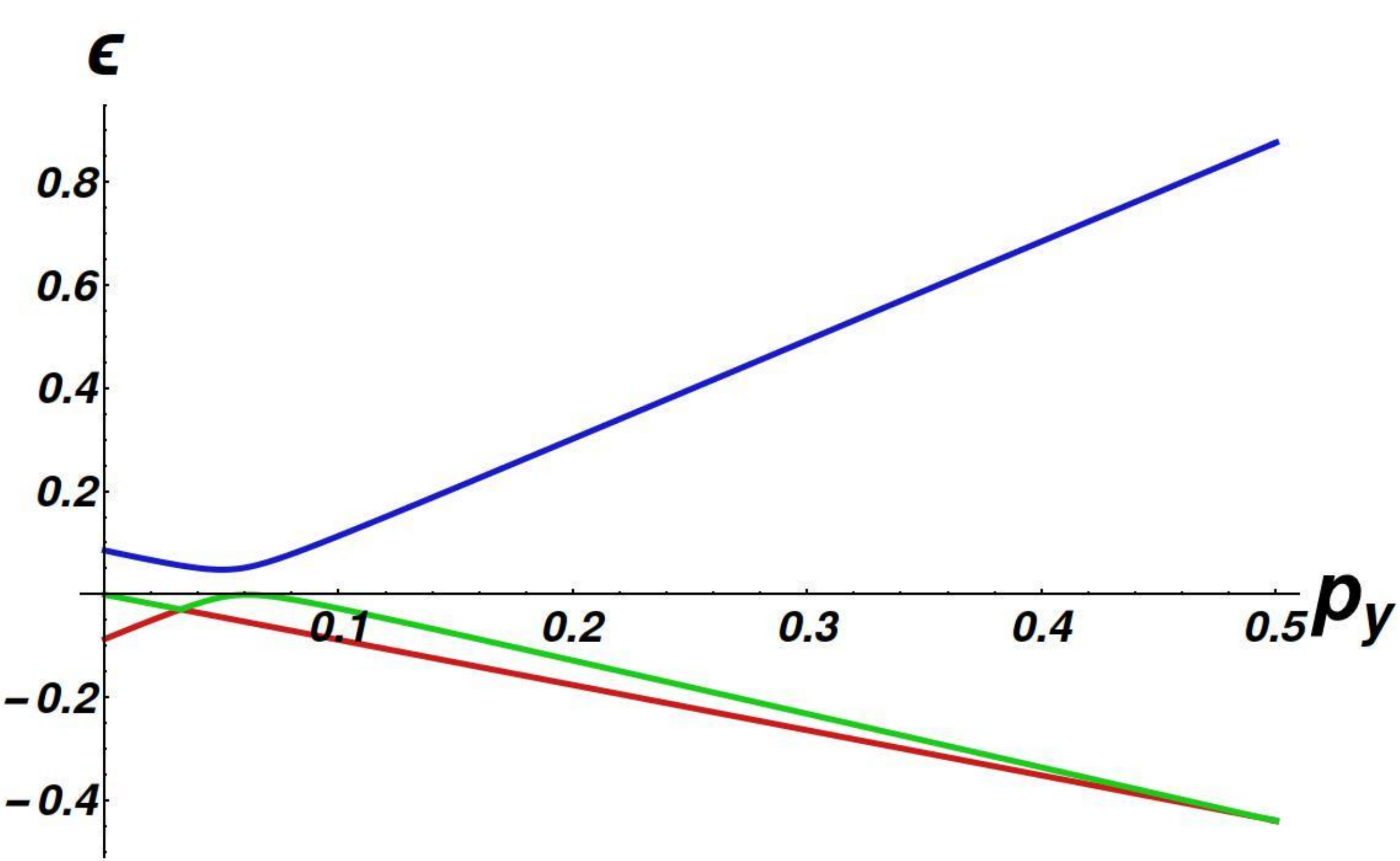}
\caption{Case 3b: $l^+_1 \neq 0$, $n^- = 0$, $f^- \neq l^-_1$. The three pictures are in increasing order
in the difference between $f^-$ and $l_1^-$.}
\label{fig:f-l1-efeect}
\end{figure}

Case 3b\label{case:3b} ($l^+_1 \neq 0$, $n^- = 0$, $f^- \neq l^-_1$):
Here, the top and middle bands touch each other linearly at two points (similar to
case 3a), while 
the middle and bottom bands also touch each other linearly at two points. 
Contrasting this with case 3a, we see that the effect of 
$f^- \neq l^-_1$ is to produce two Dirac cones when there was
only one two-fold degeneracy before. This tells us that the two-fold degeneracy
in case 3a is not a standard Dirac cone. \cite{case3a_footnote}
\SPhide{Add footnote of 1 ($\theta'_{\mathbf{p}}$), 1 ($H$) and 0(t), 2(m), 2(b).}
For the top two bands, the line connecting the degeneracy point is completely flat. 
As the difference between $f^-$ and $l_1^-$ becomes larger, 
then one of the two Dirac cones goes away rather quickly (see Fig.\ref{fig:f-l1-efeect}).

\begin{figure}
\centering
\includegraphics[width=0.7\linewidth]{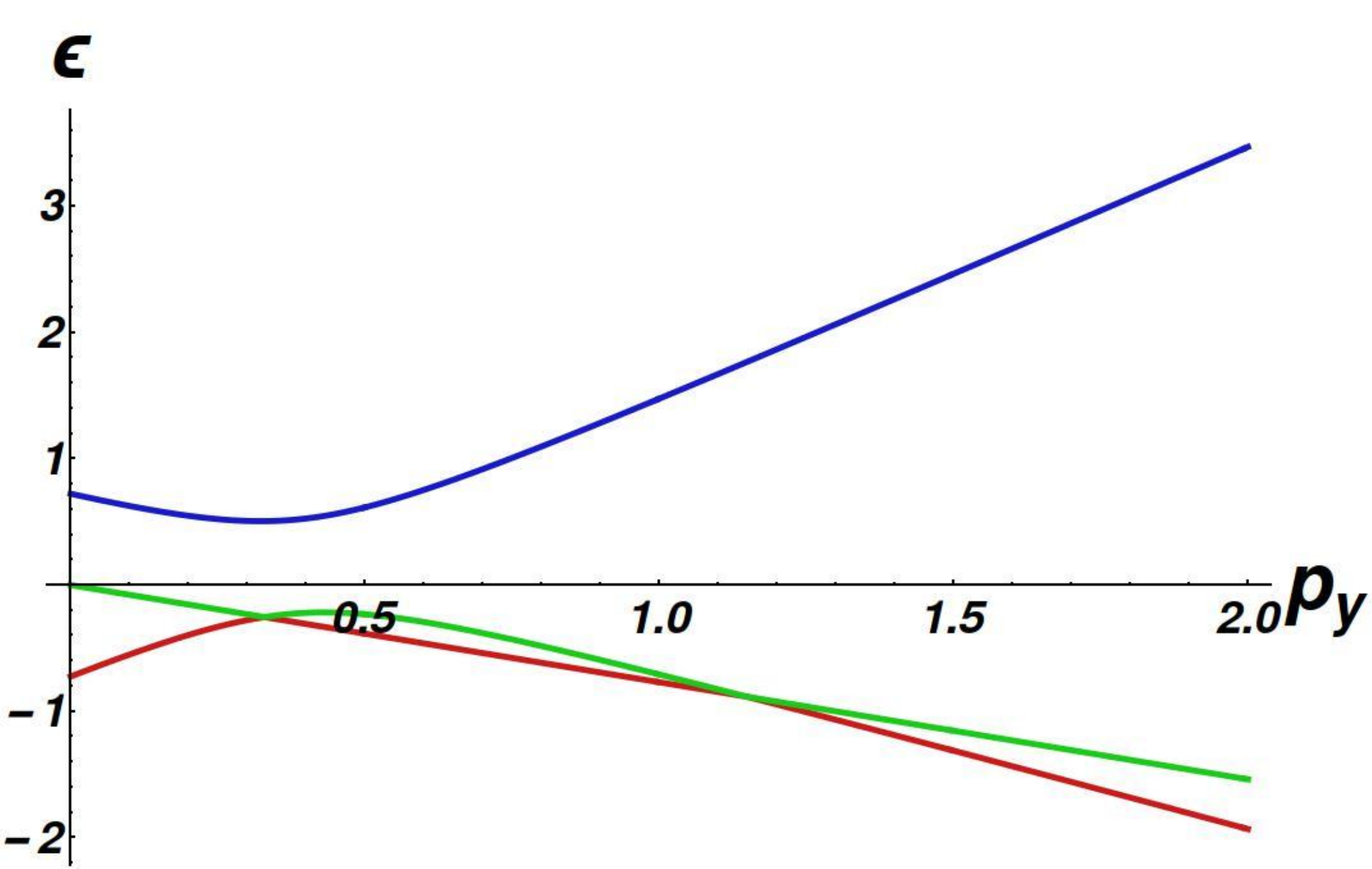}
\includegraphics[width=0.7\linewidth]{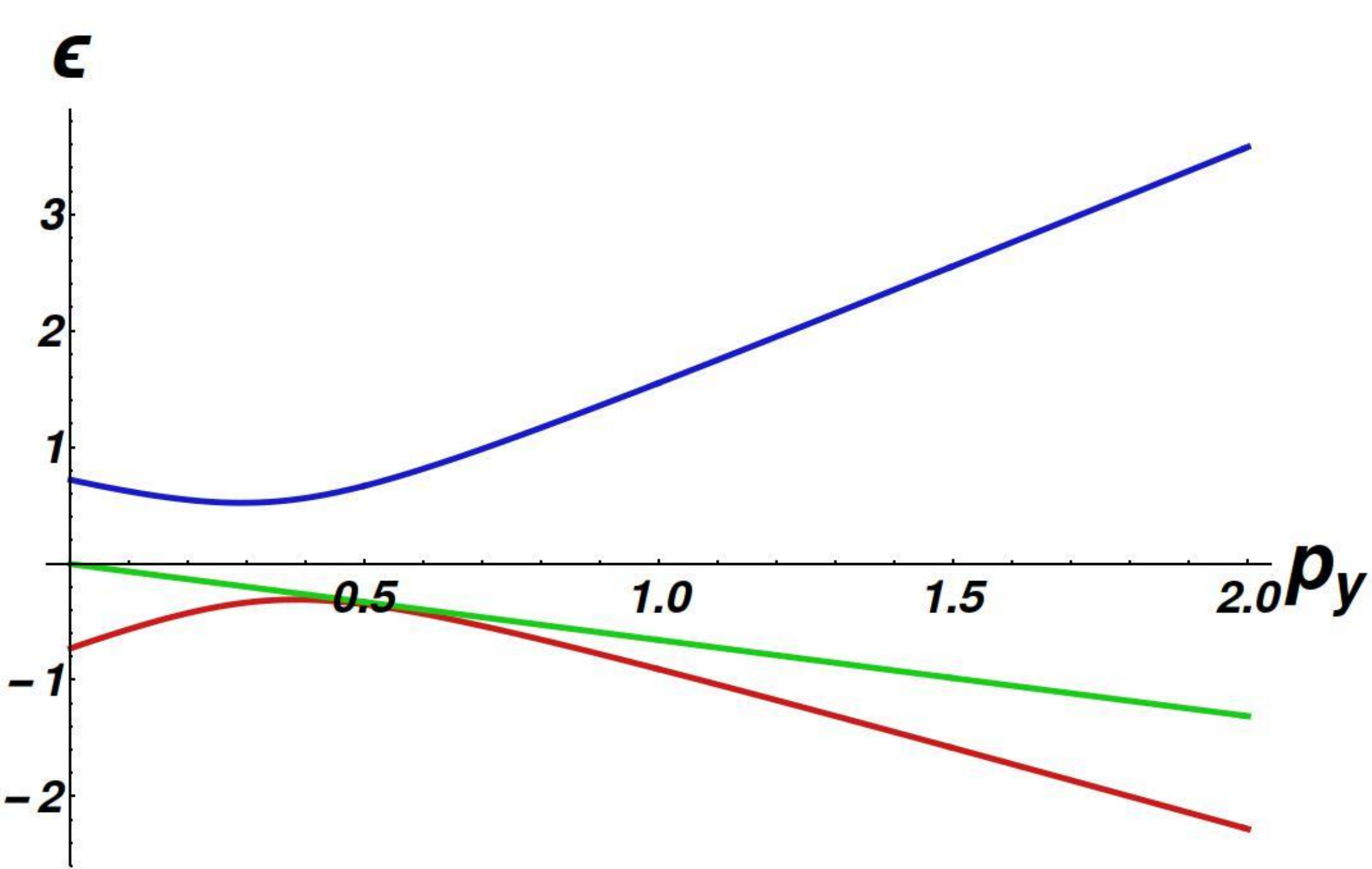}
\includegraphics[width=0.7\linewidth]{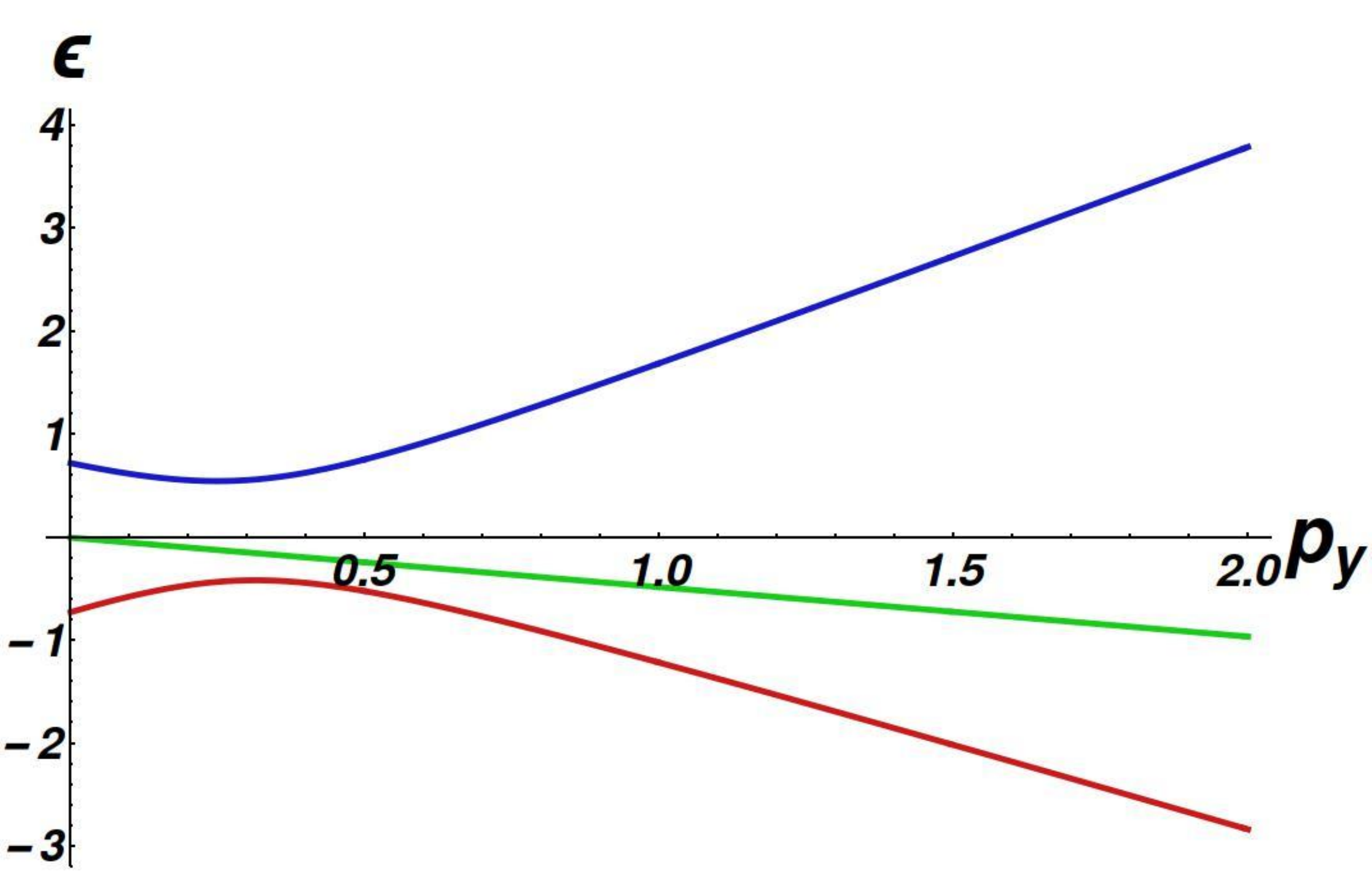}
\caption{Case 4 ($l^+_1 \neq 0$, $n^- \neq 0$):
First picture is for $n^-<n^-_{cric}$. Second picture is for $n^-=n^-_{cric}$. 
The third plot is when $n^->n^-_{cric}$ and bottom two band gap out.}
\label{fig:l8pre}
\end{figure}

Case 4 \label{case:4} ($l^+_1 \neq 0$, $n^- \neq 0$):
The diagonal momentum dependent term $n^- \tau^3 \otimes \Lambda^8$ 
comes from same sublattice hoppings. The effect of this 
is shown in Fig.\ref{fig:l8pre}.
We see that for strong enough $n^-$ the three band problem becomes an
effective two band problem with two Dirac cones connecting the top and middle bands, while the bottom band is independent. For small $n^-$ on the other hand,
 there are two Dirac cones connecting middle and bottom bands as well.
 $n^-_{cric}$ depends on other parameters in a detailed way which we do not concern ourselves with.

\begin{figure}
\centering
\includegraphics[width=0.7\linewidth]{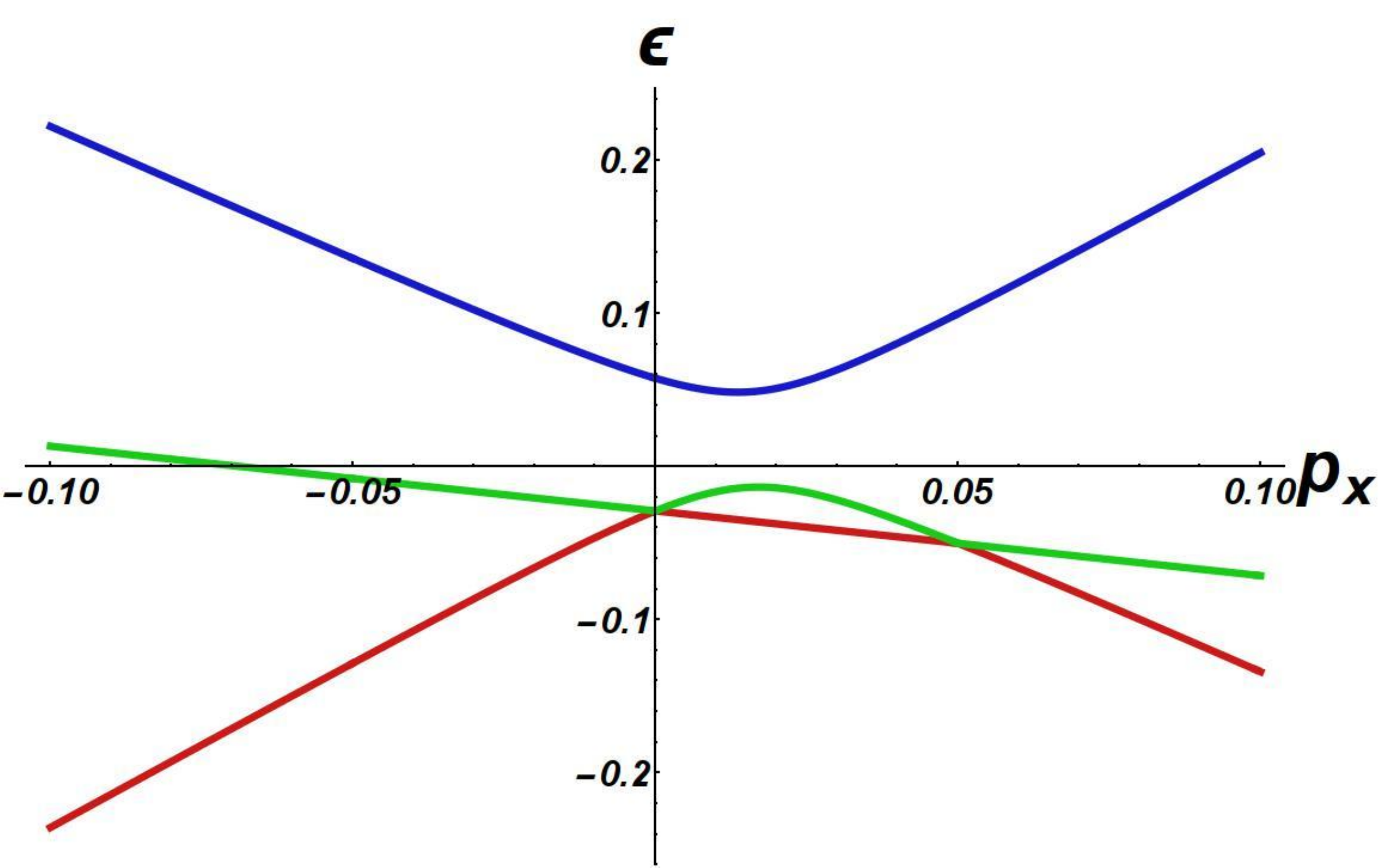}
\caption{Case 5 ($n^+ \neq 0$). This figure represents the case where $n^+<0$. If we change to $n^+>0$ then the bottom band becomes standalone.}
\label{fig:l8np}
\end{figure}

Case 5 \label{case:5} ($n^+ \neq 0$):
When the diagonal momentum independent term $n^+ \tau^0 \otimes \Lambda^8$
becomes non-zero, it effectively renders the Hamiltonian into a 
sum of two-fold bands and a standalone band. This effect is shown
in Fig. \ref{fig:l8np}. This may be thought of as similar to the effect of a (sublattice) mass term in Graphene due to
different chemical potentials on the two sublattices. However,
$n^+ \tau^0 \otimes \Lambda^8$ is allowed by $\mathcal{C}_2$ symmetry,
whereas different chemical potentials in Graphene is forbidden
by $\mathcal{C}_2$ symmetry.

The effect of $g^-$ and $m^-_2$ is innocuous in the preceding cases, and the above
categorization goes through. 
We reiterate that all purely point-degenerate band touchings above
have expected Berry phase windings
that are $\pm \pi$, or their multiples as in
the exception of Case 3a which is similar to bigraphene.\cite{Mikitik_Sharlai2008}

\section{Lattice Model}
\label{sec:lattice}
\begin{figure}
    \centering
    \includegraphics[scale=0.6]{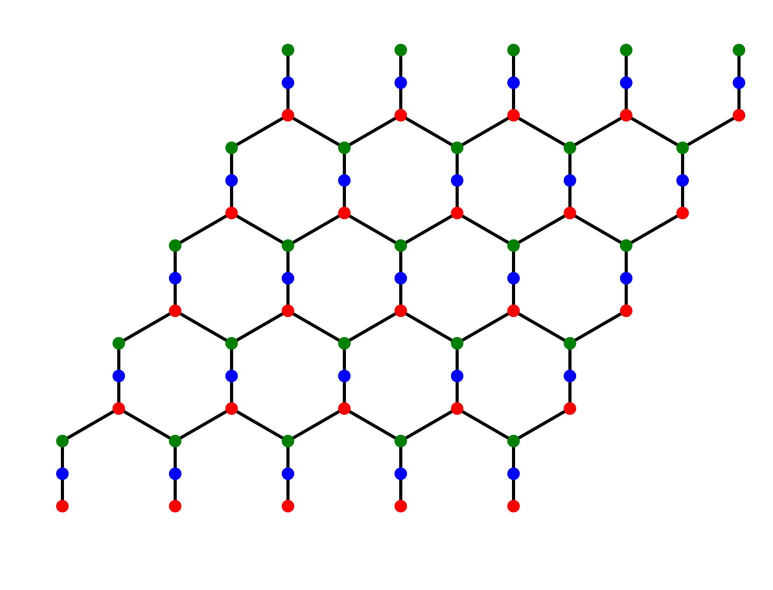}
    \caption{The Graphene-like lattice
on which fermions live is shown above. Red points are $a$ sublattice sites, green  points 
are $b$ sublattice points, blue points are $c$ sublattice sites. The underlying Bravais lattice
is still triangular same as Graphene.}
    \label{fig:lattice}
\end{figure}
In this section, we write down a Graphene-like lattice fermion model motivated
by the plausibility of realizing
the continuum Hamiltonians described in previous
sections, Eq. \ref{eq:3bandcontham} and \ref{eq:final_3bandcontham}, in some real-world material. 
The lattice that we consider is shown in
Fig. \ref{fig:lattice}. It is chosen to be very similar to the 
Graphene lattice with an extra lattice site in the
middle of vertical bonds.  
On this lattice, apart from the conventional hopping matrix elements as in 
Graphene between $a$ to $b$ sublattices, we also include hopping matrix elements
between $a$ and $c$ sublattices as well as  $b$ and $c$ sublattices. 
We note that the geometric
distance between $a$-$c$ and between $b$-$c$ inside the same unit cell is smaller than between $a$-$b$. For inter-unit cell hoppings, the situation is the opposite. So generically these hopping strengths are not equal. This lattice model can either be thought of as a planar model,
or a two-layer model where one of the layers is hexagonal and the other layer is triangular. Thus, the $c$ sublattice sites
are not symmetry related to the $a,b$ sublattice sites. 
The $a,b$ sublattice sites may be related by either a $\mathcal{C}_2$ rotation symmetry
with any $c$ site as the rotation center, or by a reflection around an axis form by joining
a horizontal row of $c$ sites.

The hopping Hamiltonian that we consider on this graphene-like lattice going by
the above symmetry considerations is given by Eq. \ref{eq:lattice_hopping_model},
where $n_1,n_2$ are unit-cell indices using the primitive lattice vectors of the underlying
triangular lattice. 
Here, we are not considering staggered potentials,
whose effect is discussed in section \ref{subsec:various_cases}
(in particular case 5: $n^+ \tau^0 \otimes \Lambda^8$; 
a term like $\tau^0 \otimes \Lambda^3$ is ruled out 
by $\mathcal{C}_2$). 

\onecolumngrid

\begin{figure}
    \centering
\includegraphics[width=0.45\linewidth]{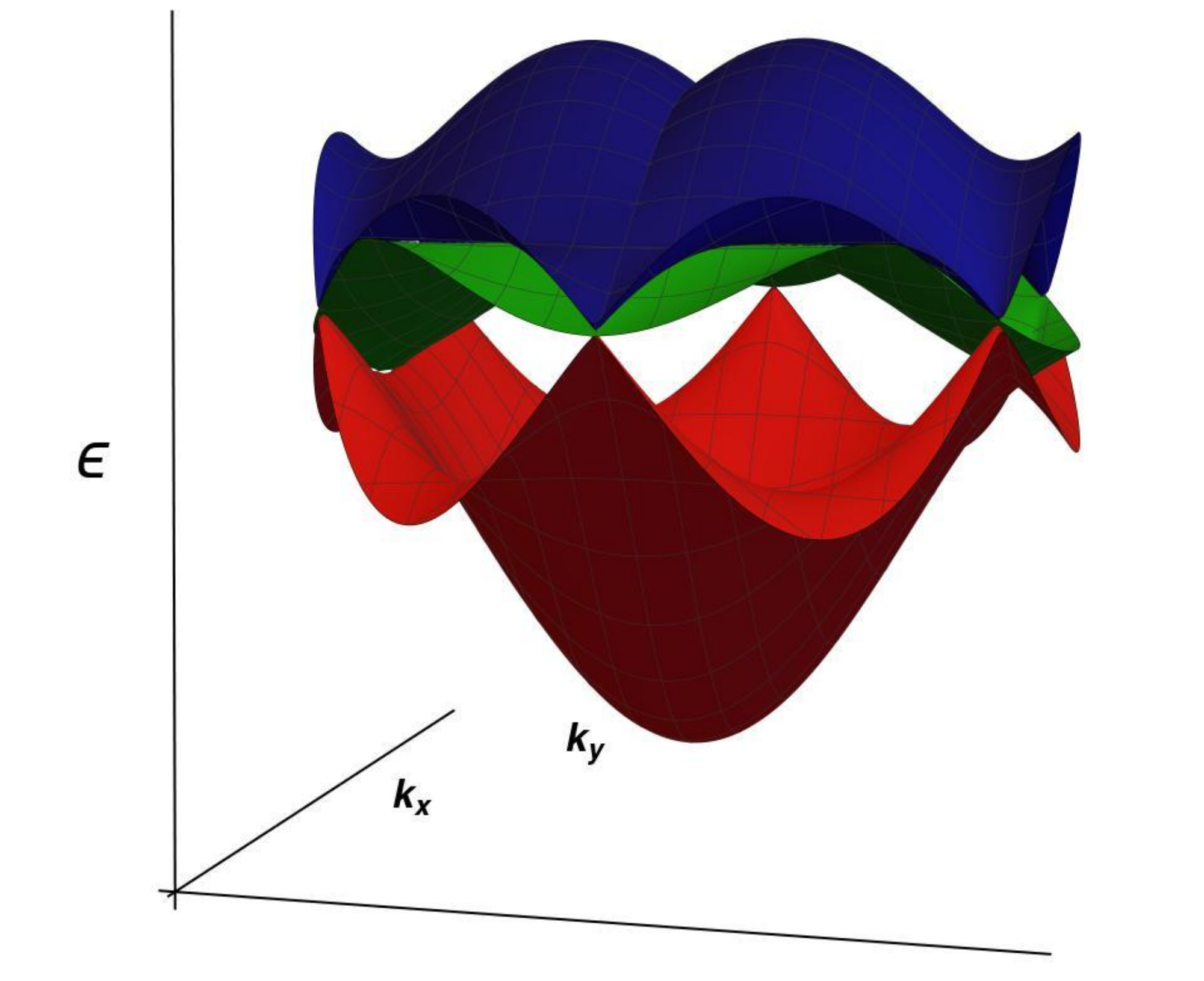}
\includegraphics[width=0.45\linewidth]{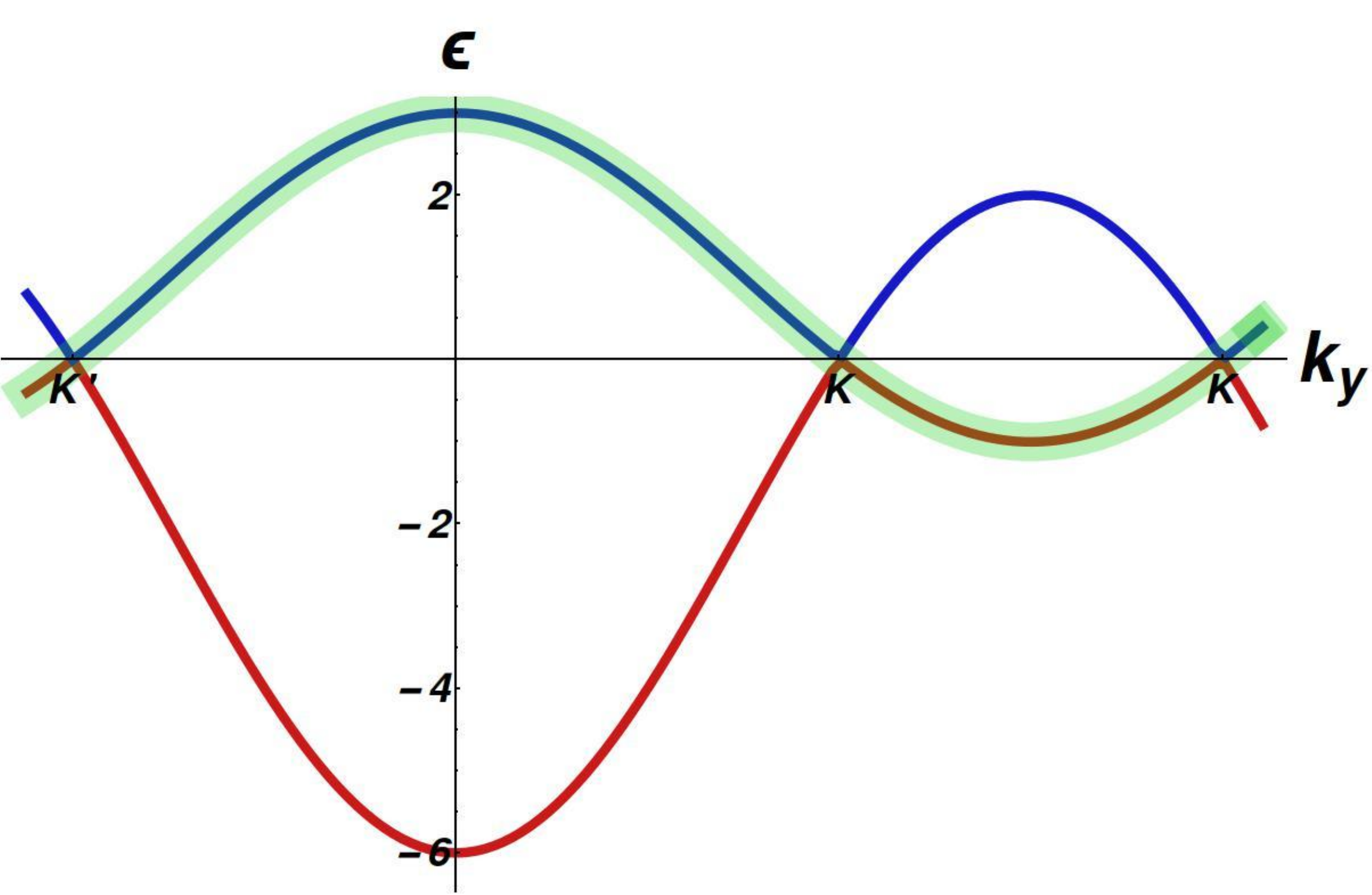}
\caption{(Left) Dispersion of the Graphene-like lattice
Hamiltonian for $\delta t_0=0$ and $\delta t_1=0$ over the full Brillouin zone.
Near the valley points $\vec{K},\vec{K}'$, they reproduce the continuum band structure
of $H_\mu^{\text{3A}}$ as in Fig. \ref{fig:contband}. (Right) 
A cut along $p_y=0$ is shown to better display the non-contractible loop
on which the two-fold line degeneracy lives.}
\label{fig:3bndz}
\end{figure}

\begin{align}
\mathcal{H} & = \sum_{n_1,n_2} \mathcal{H}_{ab}+\mathcal{H}_{ac}+\mathcal{H}_{bc} \\
\mathcal{H}_{ab} & = 
- t ~\hat{c}^\dagger_{(n_1,n_2),a} \left( \hat{c}_{(n_1,n_2),b}+\hat{c}_{(n_1,n_2-1),b}
+\hat{c}_{(n_1+1,n_2-1),b} \right) + \text{h.c.} \nonumber \\
\mathcal{H}_{ac} &= -(t+\delta t_0)\hat{c}^\dagger_{(n_1,n_2),a} \hat{c}_{(n_1,n_2),c}
-(t+\delta t_1)\hat{c}^\dagger_{(n_1,n_2),a} \left( \hat{c}_{(n_1,n_2-1),c} + \hat{c}_{(n_1+1,n_2-1),c} \right) + \text{h.c.} \nonumber \\
\mathcal{H}_{bc} &= -(t+\delta t_0)\hat{c}^\dagger_{(n_1,n_2),b} \hat{c}_{(n_1,n_2),c}
-(t+\delta t_1)\hat{c}^\dagger_{(n_1,n_2),b} \left( \hat{c}_{(n_1,n_2+1),c} + \hat{c}_{(n_1-1,n_2+1),c} \right) + \text{h.c.}
\label{eq:lattice_hopping_model}
\end{align}

\twocolumngrid

This lattice Hamiltonian clearly reproduces the continuum $H_\mu^{\text{3A}}$
band structure near the valleys $\vec{K},\vec{K}'$ when the ``deformations"
$\delta t_0$ and $\delta t_1$ are zero as may be seen by expanding to leading
order near these points in the zone. Essentially they are three copies
of Graphene hoppings with the same strength. The dispersion over the full Brillouin zone 
for this choice of parameters is shown in the left panel of
FIG.\ref{fig:3bndz}.
The two-fold line degeneracies that connect the two valleys $\vec{K},\vec{K}'$
form a \emph{non-contractible loop} in the Brillouin zone. This is shown in the right panel
of FIG.\ref{fig:3bndz}. 

For generic deformations, $\delta t_0 \neq0$, $\delta t_1\neq 0$, the two-fold line degeneracies go away.
Then, the continuum Hamiltonians near the two valleys look like
\onecolumngrid

\begin{equation}
    \mathcal{H}^{\vec{K}}(\vec{p})=\frac{\sqrt{3}}{2}(1+\delta t_1)
           \left(\begin{matrix} 
            0 & \frac{1}{(1+\delta t_1)}(p_x-i p_y) & \frac{2(\delta t_1-\delta t_0)}{\sqrt{3}(1+\delta t_1)}+(p_x - i p_y) \\
            \frac{1}{(1+\delta t_1)}(p_x + i p_y) & 0 & \frac{2(\delta t_1-\delta t_0)}{\sqrt{3}(1+\delta t_1)}+ (p_x + i p_y)\\
            \frac{2(\delta t_1-\delta t_0)}{\sqrt{3}(1+\delta t_1)}+(p_x + i p_y) & \frac{2(\delta t_1-\delta t_0)}{\sqrt{3}(1+\delta t_1)}+(p_x - i p_y) & 0
           \end{matrix}\right)
        \label{eq:3bandhamcontgen}
\end{equation}
\begin{equation}
    \mathcal{H}^{\vec{K}'}(\vec{p})=\frac{\sqrt{3}}{2}(1+\delta t_1)
           \left(\begin{matrix} 
            0 & \frac{1}{(1+\delta t_1)}(-p_x-i p_y) & \frac{2(\delta t_1-\delta t_0)}{\sqrt{3}(1+\delta t_1)}+(-p_x - i p_y) \\
            \frac{1}{(1+\delta t_1)}(-p_x + i p_y) & 0 & \frac{2(\delta t_1-\delta t_0)}{\sqrt{3}(1+\delta t_1)}+ (-p_x + i p_y)\\
            \frac{2(\delta t_1-\delta t_0)}{\sqrt{3}(1+\delta t_1)}+(-p_x + i p_y) & \frac{2(\delta t_1-\delta t_0)}{\sqrt{3}(1+\delta t_1)}+(-p_x - i p_y) & 0
           \end{matrix}\right)
        \label{eq:3bandhamcontgenpr}
\end{equation}
\twocolumngrid

This deformed Hamiltonian is in the form found in Sec. \ref{sec:su3}, Eq. \ref{eq:final_3bandcontham}
consistent with all our symmetry considerations. 
We reiterate that for our lattice hopping model,
we have $n^+ = n^- = f^+ = 0$ in the notation of Sec. \ref{sec:su3}.
The reasons are as follows: a) $n^+$ and $n^-$ are zero because there
are no staggered chemical potentials and no inter-cell same sublattice hoppings.
b) $f^+ = 0$ because we are measuring deformations in hopping with respect
to $ab$ hoppings within the unit cell.
Secondly, $l^+_1 \neq 0$ is the way we chose to organize the deformations
due to $\delta t_0$, $\delta t_1$ as shown in Eq. \ref{eq:3bandhamcontgen} and \ref{eq:3bandhamcontgenpr}.
Therefore, the rest of the Hamiltonian is the undeformed case, which
makes $f^- = g^-$ and $m_2^- = l_1^-$. In fact, $f^- = g^-$ is precisely what
happens in Graphene.

\begin{figure}
    \centering
    \includegraphics[width=0.8\linewidth]{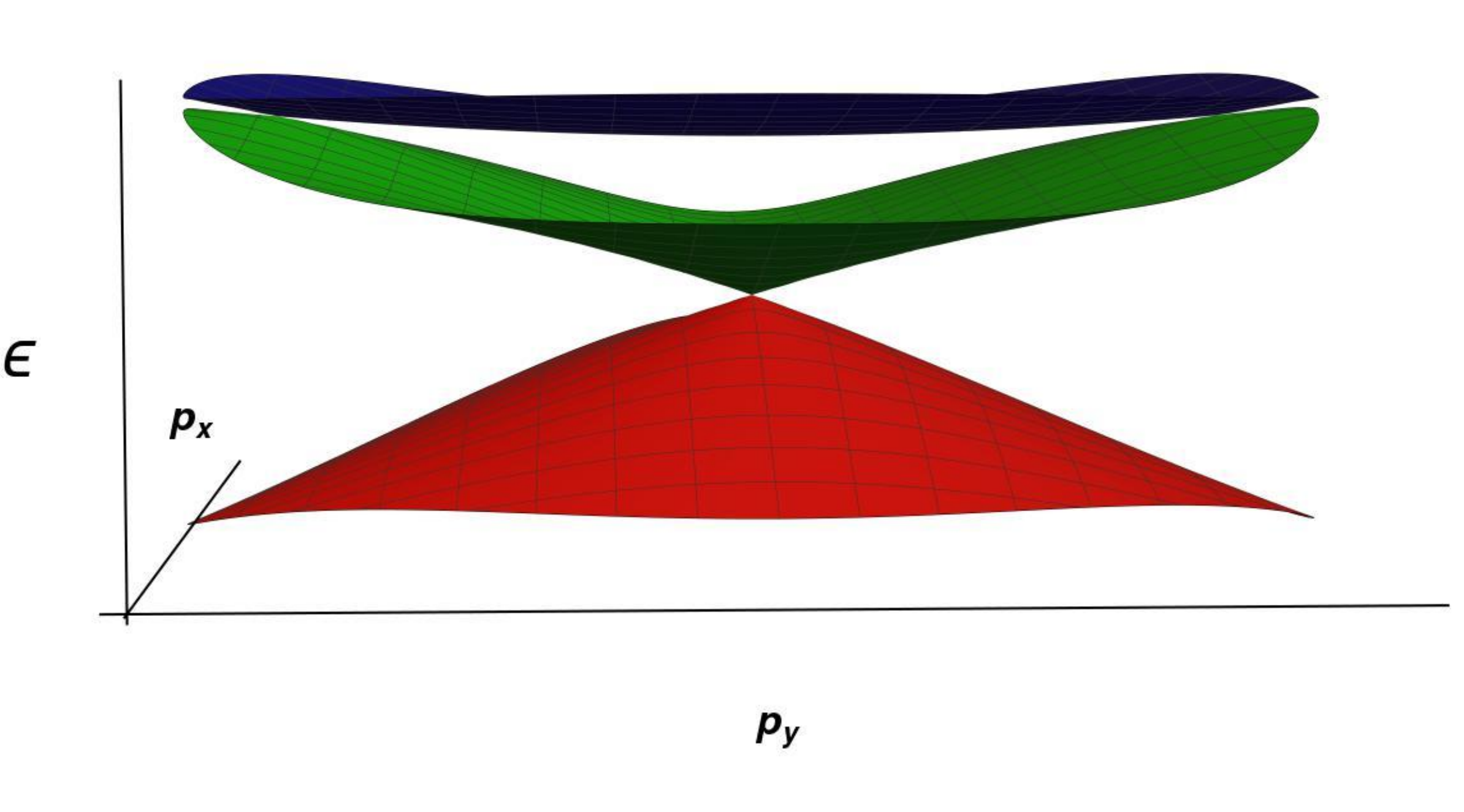}
    \caption{For $\delta t_0=0.5$ and $\delta t_1=-0.25$ the dispersion has this behavior where the bottom band and the middle band has one Dirac point and the middle band and the top band has two Dirac point.}
    \label{fig:3BDrnz}
\end{figure}

The band structure for this generic case ($\delta t_0 \neq \delta t_1 \neq 0$)
near the valleys is shown in
Fig. \ref{fig:3BDrnz}. We find that the bottom two bands touch linearly as in a Dirac cone, while
the middle and the top band involve two Dirac cones.  The dispersion along the line
connecting the two Dirac cones is rather flat. This is basically the lattice realization
of case 3 in Sec. \ref{sec:su3} where the line connecting the two Dirac cones is completely flat.
The lack of complete flatness for the lattice case is due to subleading terms.

\begin{figure}
\includegraphics[width=0.7\linewidth]{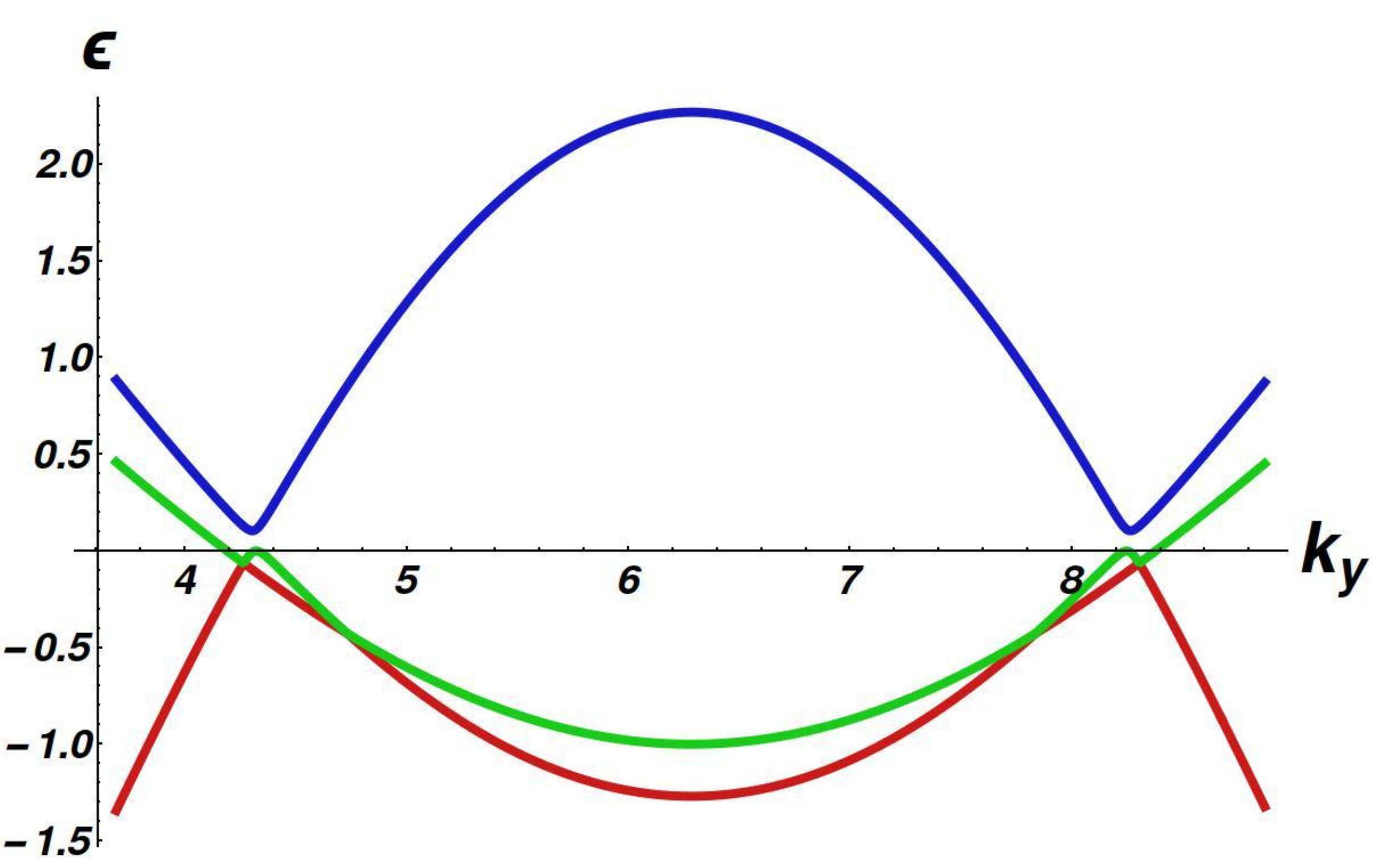}
\includegraphics[width=0.7\linewidth]{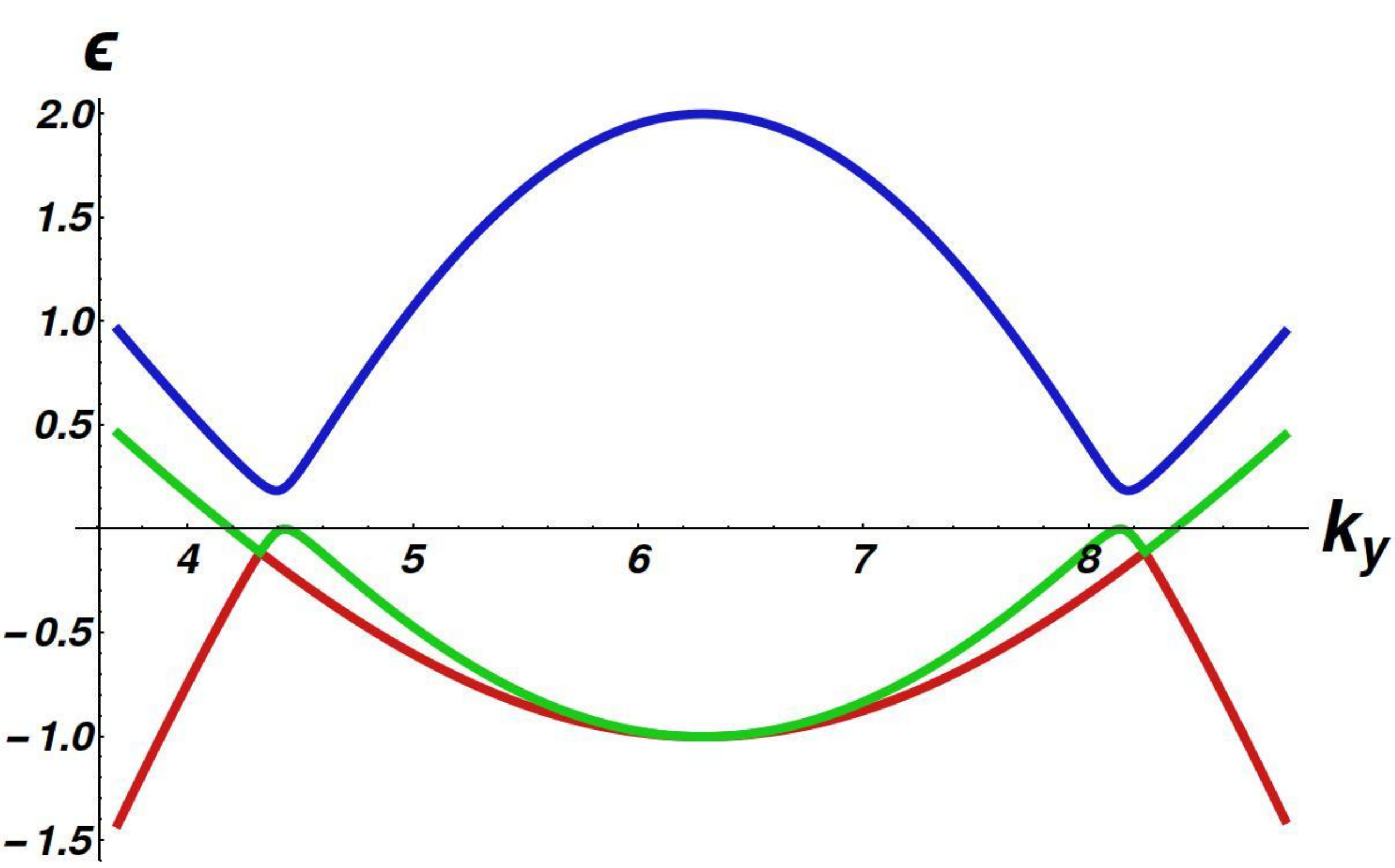}
\includegraphics[width=0.7\linewidth]{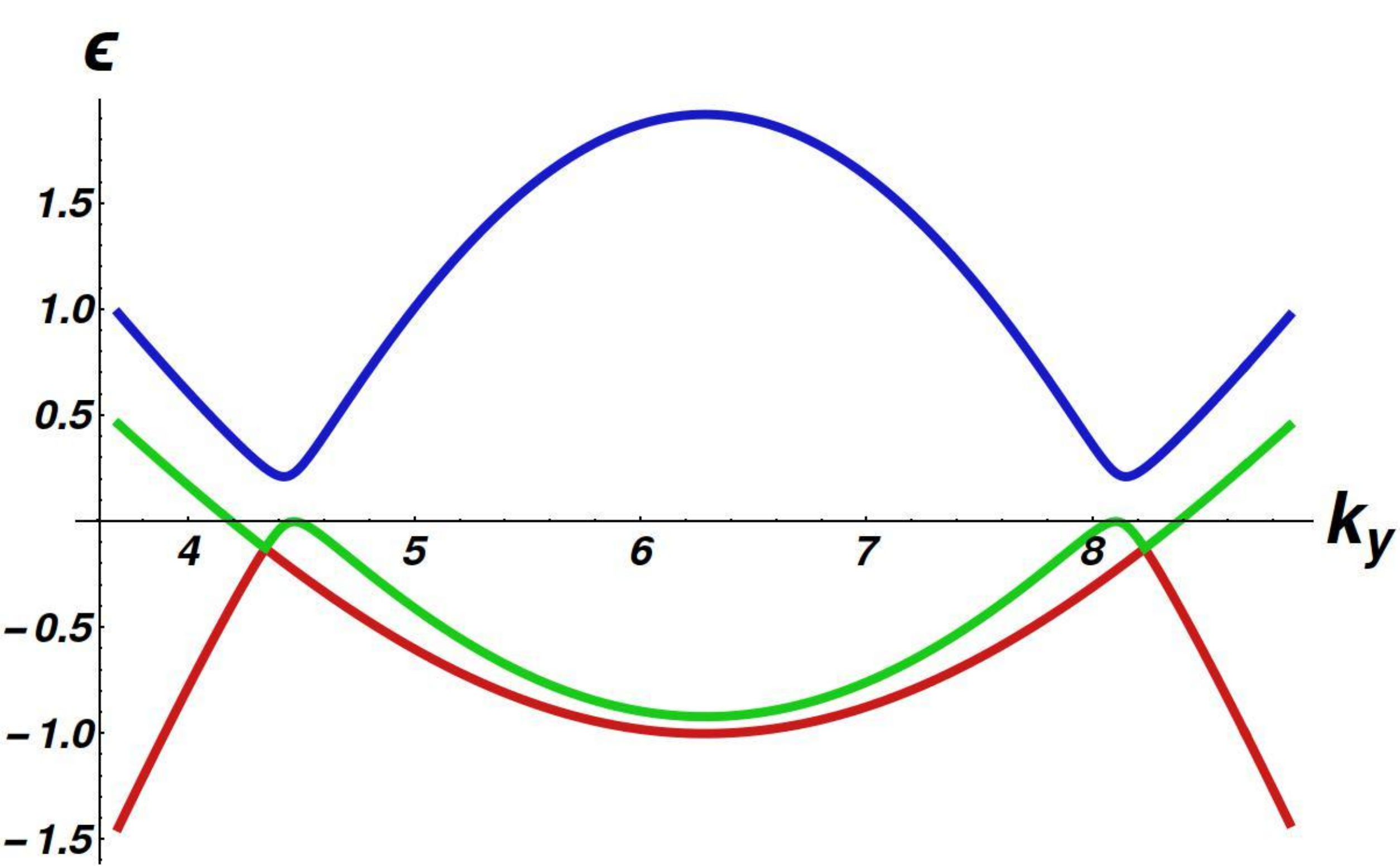}
\caption{
Three figures are for $\delta t_0=0.5$ and $\delta t_1=0.35.0.25,0.22$ respectively. As we see here the extra Dirac point travels and gets finally gapped out.}
\label{fig:ditrav}
\end{figure}

\begin{figure}
    \centering
    \includegraphics[width=0.7\linewidth]{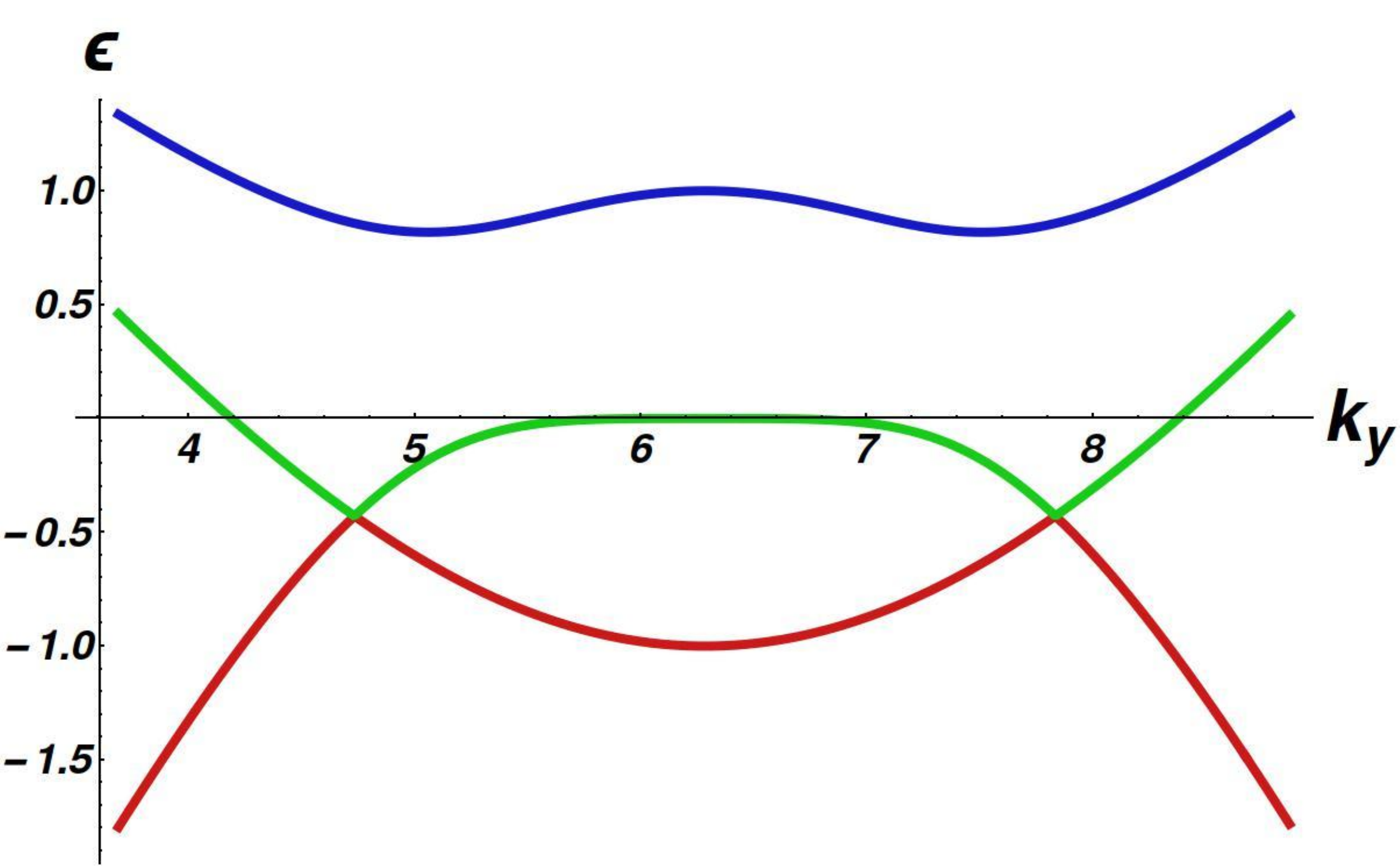}
    \caption{For $\delta t_0=0.5$ and $\delta t_1=-0.25$ the dispersion has this behavior along $k_y=0$ line}
    \label{fig:py0}
\end{figure}

For the middle and bottom band, there can in fact be another Dirac cone apart from
the one mentioned above, which can annihilate with a similar counterpart from the
other valley as we tune $\delta t_1$ for a given $\delta t_0$ as shown in Fig. \ref{fig:ditrav}.
This happens when $\delta t_1$ and $\delta t_0$ are of the same sign. When their signs
are opposite, they are always annihilated as shown in Fig. \ref{fig:py0}. In our Graphene-like lattice, we expect the
opposite sign case to be the physical case
if we assume that hopping strength decreases with distance.

We also mention a fine-tuned case of deformation when $\delta t_0 = \delta t_1 \neq 0$.
The corresponding band structure near the valleys is included
in case 2 in Sec. \ref{sec:su3} where we again have a three-fold degeneracy
without two-fold line degeneracies. A final comment on breaking the $\mathcal{C}_2$ symmetry
and thereby opening gaps near the Dirac cones: this makes the gapped bands topologically
trivial bands with zero Chern number just like Graphene.

\subsection{Effect of magnetic field}
\label{subsec:magnetic_field}

To conclude this section, we quickly discuss the Landau levels of our lattice model in the presence
of a perpendicular magnetic field which is mainly a numerical study. 
The Landau level structure of Graphene and its multi-layer variants\cite{Novoselov_etal2006}
have received attention  due to their different quantization properties than the
$2d$ electron gas. This motivates us to discuss the Landau levels in our case because
of the presence of the non-standard ``three-band" Dirac cone structures as discussed previously.

\begin{figure}[H]
    \centering
\includegraphics[width=0.8\linewidth]{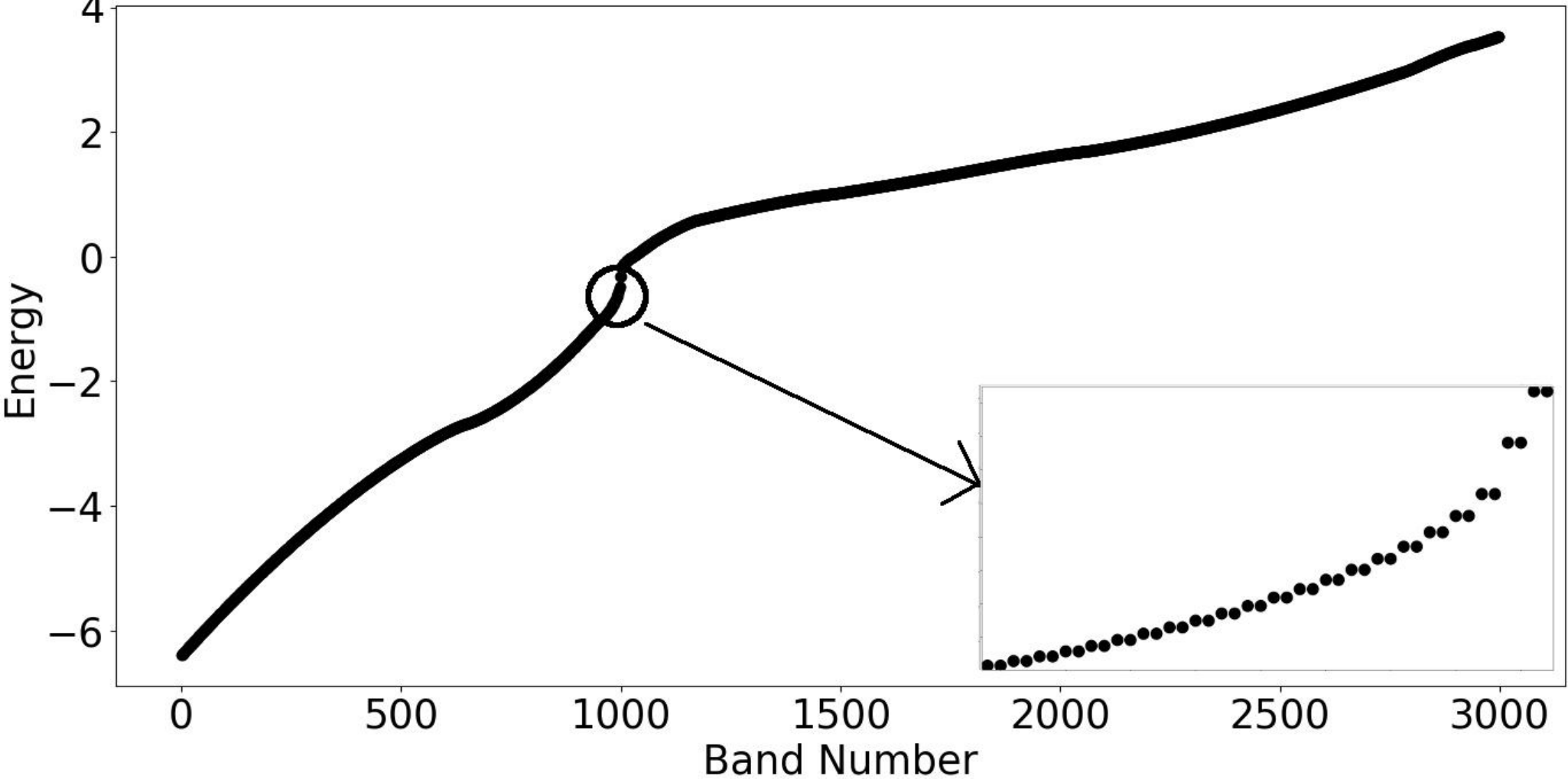}
\caption{Here we present the spectrum for $\delta t_0=0.5$,
$\delta t_1=-0.25$ where $p=1,q=1000$.
In the inset it shows that for the bottom band the spectrum is $\sim\sqrt{n}$ and they are doubly degenerate.}\label{fig:hofbut_largeq}
\end{figure}
We start by showing our numerical computation of the Landau levels in the Hofstadter limit
for our model with hopping deformations
for a very large $q=1000$ in Fig. \ref{fig:hofbut_largeq}. We can identify
regions in this diagram that are linear where the underlying band is dominantly quadratic (e.g.
near the very bottom and top of the three bands),
and regions that are square-root like where the underlying band is dominantly linear (e.g.
near Dirac cones). These features are marked in Fig. \ref{fig:hofbut_largeq}. In the region
where the top band and middle band touch with two separate Dirac cones, 
we find that the behavior
is neither linear nor square-root like. Numerically fitting this behavior gave us a
power close to $7/9$.

Going by the usual steps at the continuum level, we run into difficulty.
E.g. for the case of $H^{\text{3A}}_K$, we arrive at a Landau level Hamiltonian
that is proportional to
$\left(\begin{matrix} 
            0 & \hat{a}^\dagger & \hat{a}^\dagger \\
            \hat{a} & 0 & \hat{a}\\
            \hat{a} & \hat{a}^\dagger & 0
           \end{matrix}\right)$
where $\hat{a}=\frac{l_\text{B}^2}{2\hbar}\mathbb{P}_+$ and $\hat{a}^\dagger=\frac{l_\text{B}^2}{2\hbar}\mathbb{P}_-$,
$\mathbb{P}_+=(p_x+eBy)+i p_y$ and $\mathbb{P}_-=(p_x+eBy)-i p_y$
and $\left[ \mathbb{P}_+,\mathbb{P}_- \right]= \frac{2\hbar}{l_\text{B}^2}$.
It is not clear how to derive the Landau level quantization starting with this.
One may however attempt to do a semi-classical analysis \cite{Fuchs_etal2010, Ozerin_Falkovsky2012} when $\delta t_0>0$ and $\delta t_1<0$ (non-zero deformation of hopping) such that there are well defined closed electron orbits. 
This is asymptotically valid for $n \gg 1$. For regions where the band structure is quadratic/linear,
this formula will yield the usual behaviors of $n$ and $\sqrt{n}$ respectively as also seen in our numerical computations
(Fig. \ref{fig:hofbut_largeq}). 
The numerical results have been obtained by diagonalizing the Hofstadter 
problem for very large $q$, equivalently for very small magnetic fields.

\begin{figure}
    \centering
    \includegraphics[width=0.7\linewidth]{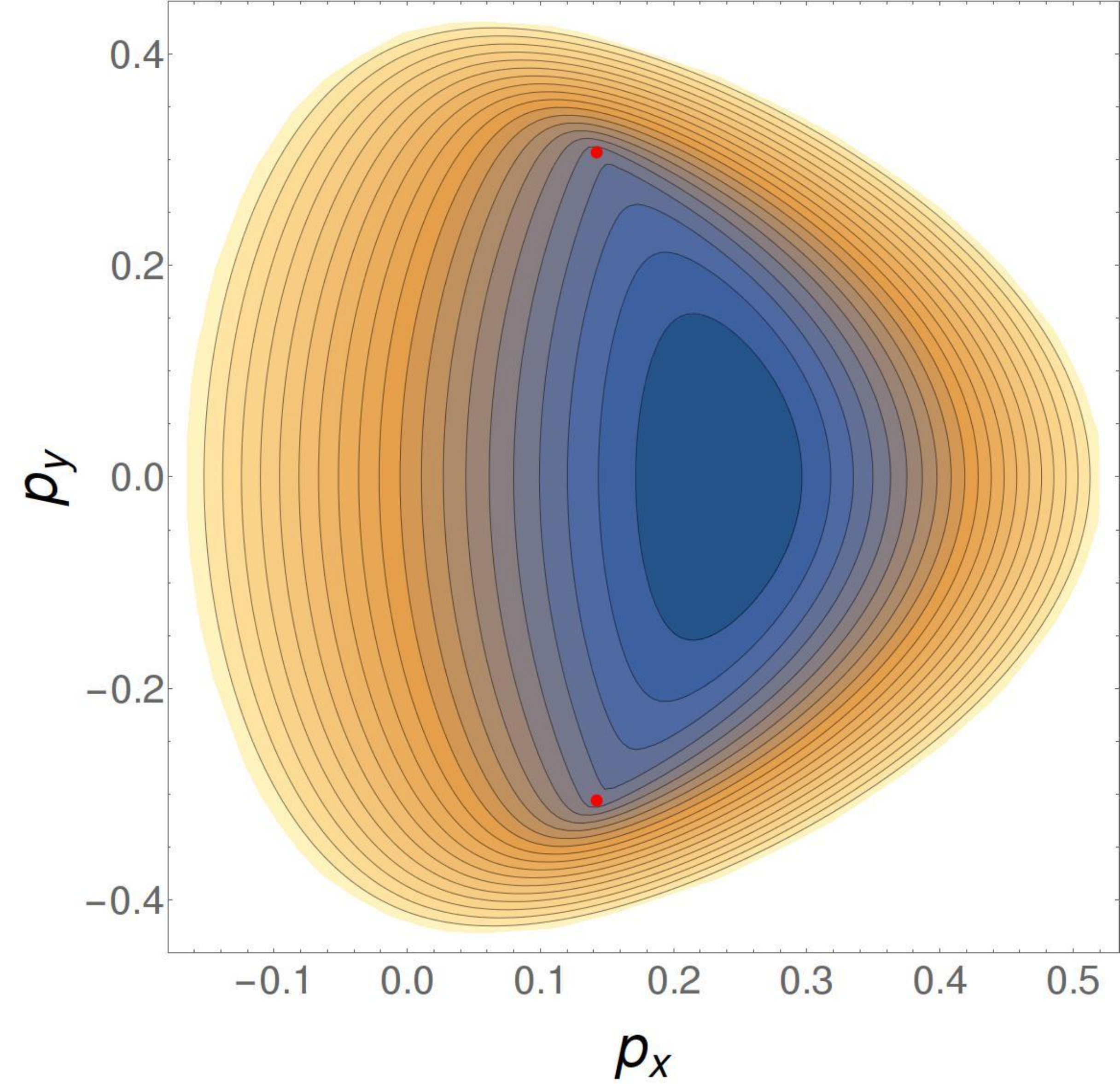}
    \caption{Orbit of the electron in the top band near the two same energy band degeneracy for the top band. The red points signify the location of the Dirac points.}
    \label{fig:orbit_shape}
\end{figure}

Near the unusual two Dirac cone structure between
the top and middle band, the orbits have non-standard
shapes as shown in Fig. \ref{fig:orbit_shape} with the left side scaling linearly in energy, while
the right side scaling quadratically in energy. Crudely estimating the area of such orbits leads
to the conclusion that Landau level behavior will be somewhere in between $n$ (coming from
quadratic scaling part of the orbit) and $\sqrt{n}$ (coming from the linear scaling part of the orbit).
However, it does not yield a neat power-law, but since the semi-classical analysis is applicable
only in the $n \gg 1$ limit, we may guess that the numerical observation of $\sim 7/9$ exponent
is due to the quadratic scaling part of the orbit eventually dominating the orbited area.

\begin{figure}
    \includegraphics[width=0.9\linewidth]{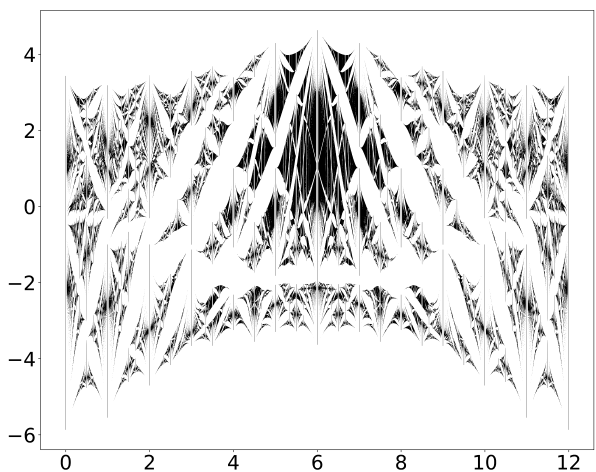}
    \caption{Hofstadter Butterfly for our model lattice with $\delta t_0=0.1$ and $\delta t_1=-0.1$. Here on $x$-axis represents the flux in units of quantum flux enclosed by the unitcell and $y$-axis represents the Energy.}
    \label{fig:hofbut}
\end{figure}

We finally show the full Hofstadter butterfly spectrum for our lattice model
in Fig. \ref{fig:hofbut} for completeness.
Here we can identify a few features of our lattice model:
1) the Hofstadter butterfly repeats after 12 quantum flux per unitcell. This is due to the fact that
in our Graphene-like lattice model, the smallest area covered by the
hopping is not the hexagonal plaquette, but $\frac{1}{12}^{\text{th}}$ of it.
2) There is no particle hole symmetry.
3) For $\frac{1}{2}$ flux quanta per smallest area, the model still has time reversal symmetry and thus there is no gap.

\vspace*{10pt}

\section{Conclusion and Outlook}
\label{sec:conclusion}

In summary, the central result of this paper is
the continuum Hamiltonian $H^{\text{3A}}_K$ (Eq. \ref{eq:3bandcontham})
and its eigensystem (Eq. \ref{eq:3band_eigensystem})
that we wrote down as a three-band generalization of the $2d$ Dirac Hamiltonian.
We were led to consider them in order to arrive at a beyond-Dirac-like or non-$SU(2)$ geometric
phase structure in two dimensions as our primary motivation (Sec. \ref{sec:intro}). 
We exposed the geometric phase structure of $H^{\text{3A}}_K$ using
a triplet of indices as described in Sec. \ref{sec:3band} and summarized in
Table \ref{tab:class}. Through this table, we see how $H^{\text{3A}}_K$ contrasts
with other cases that have $SU(2)$ geometric phase structure.

Guided by the $SU(3)$ nature of $H^{\text{3A}}_K$,
we constructed in Sec. \ref{sec:su3} the general family of
continuum 2$d$ Hamiltonians (Eq. \ref{eq:final_3bandcontham})
with fermions (at a valley) in the $SU(3)$ fundamental
representation, that is allowed by time reversal $\mathcal{T}$ symmetry, 
and the space symmetries of inversion $\mathcal{C}_2$             
and reflections $\mathcal{P}_x$, $\mathcal{P}_y$.  $H^{\text{3A}}_K$
sits at a special point in this family of Hamiltonians. We further categorized
the various three-band dispersions that result from different regions in this
family of Hamiltonians (Sec. \ref{subsec:various_cases}).

In Sec. \ref{sec:lattice},
we provided a tight-binding lattice model realization of 
$H^{\text{3A}}_K$ on a Graphene-like lattice (Fig. \ref{fig:lattice}), where the three bands touch each other 
at $K$ and $K'$ when the hopping matrix elements are appropriately fine-tuned, with a 
line of two-fold degeneracy connecting $K$ and $K'$ on a
non-contractible loop in the Brillouin zone (right panel of Fig. \ref{fig:3bndz}). 
Away from the fine-tuned point, we realize various cases of Eq. \ref{eq:final_3bandcontham}.
We studied the effect of a uniform magnetic field including its Hofstadter butterfly (Fig. \ref{fig:hofbut})
and found that the Landau level quantization is different for different parts of the spectrum (Fig. \ref{fig:hofbut_largeq}).

We end the summary with a conceptual remark.
Our discussion on the geometric phase structure
of the $H^{\text{3A}}_K$ (Eq. 3) in Sec. \ref{sec:3band} 
shows that there 
is a way to construct a topological invariant in presence
of line degeneracies.
Often, geometric phases in two dimensions
are discussed by considering a closed orbit around some point  degeneracy,
chosen such that the (restricted) one-dimensional 
band structure on the orbit is \emph{gapped} throughout.
The winding number of the wavefunction's phase in this restricted dimension 
then serves as a topological invariant that characterizes
the geometric phase structure around the point degeneracy in the higher dimension\cite{Ryu_2010}.
Here, we have shown that there is a way to generalize this approach
for a three-fold point degeneracy
in two dimensions with line degeneracies, and thereby no adiabaticity
in the sense of Berry \cite{Berry1984}.
As discussed in the text (Eq. 4 and para below),
analytical continuation of the eigenvalues and eigenvectors on a
one-dimensional loop \emph{across} the line degeneracy while considering a closed orbit around
the three-fold point degeneracy (Fig. 2) is what allows for this
generalization.
This idea of analytical continuation is then used 
-- \emph{and may be used more generally in other situations} --
for an appropriate topological invariant characterization
of the geometric phase structure in the restricted dimension 
even in presence of \emph{gapless} points.
This may be considered as a new lens on the discussion of geometric phase structure
of two-dimensional band structures, and possibly in higher dimensions, e.g. 
giving a perspective
on some recent striking three-dimensional band structures 
\cite{Chang_etal2017a,Barman_etal2019} whose cuts in two dimensions
harbor three-fold point degeneracies with emanating 
two-fold line degeneracies as well.

In the future, it will be interesting to pursue the following lines of research motivated
by this paper. We have mainly explored three-band generalizations with two valleys. 
However, for three or higher bands it is
not obvious if there are generalized band structures which accommodate more than two
valleys in some interesting way. For example, in Graphene in the presence of a uniform
perpendicular magnetic field, it is known that there can be any even number of Dirac
points. \cite{Rhim_Park2012} 
Perhaps for $SU(3)$, something similar might be possible even in the absence
of magnetic fields including an odd number of valleys.
We have not paid attention to the spin quantum number in this paper.
One can study what new kind of terms can arise in the sense of Eq. \ref{eq:final_3bandcontham} in presence of spin-orbit coupling.
In the presence of more bands, can one realize higher representations of $SU(3)$ as well as other $SU(N>3)$. 
Apart from these questions of the ``band engineering" kind, there 
is the important question with regards to
the effect of interaction terms allowed by symmetries
on these band structures, as also the question regarding
the physical consequences
of such three-fold band structures
such as in measurements of
optical conductivity, \cite{Illes_Nicol2016,Kovacs_etal2017} magnetotransport 
\cite{Xu_Duan2017, Islam_Dutta2017} 
and atomic collapse \cite{Gorbar_Gusynin_Oriekhov2019}.

Finally, we ask ourselves where can we see our imagined non-interacting band structures in Nature.
Apart from the electronic structure on a possible Graphene-like lattice, perhaps other platforms
like photonic band systems \cite{Costas2001, Ray_Ghatak_Das2017}, cold atomic systems 
\cite{Lewenstein_etal2007, Tarruell_etal2012, Hu_etal2018, Hu_Zhang2018}  or 
designed lattice systems \cite{Tadjine_Allan_Delerue2016,Slot_etal2019} may be interesting
platforms to search for this. It remains to be seen if the beyond-Dirac-like geometric phase structure
that we studied in this paper can be observed in some $2d$ layered material system.

\begin{acknowledgments}
We thank G. Murthy and Luiz Henrique Santos for some discussions.
SP thanks the support of NSF grant DMR-1056536 during the initial conception stages.
SP thanks the IRCC, IIT Bombay (17IRCCSG011) for financial support,
and the hospitality of Dept. of Physics and Astronomy, University of Kentucky
where part of the work was completed.
AD thanks to the support of NSF grant DMR-1611161 and NSF grant DMR-1306897.
This research was supported in part 
by the International Center for Theoretical Sciences (ICTS) during a visit for participating in the program - The 2nd Asia Pacific Workshop on Quantum Magnetism (Code: ICTS/apfm2018/11).
\end{acknowledgments}

\bibliographystyle{apsrev}
\bibliography{biblio}

\begin{thebibliography}{57}
\expandafter\ifx\csname natexlab\endcsname\relax\def\natexlab#1{#1}\fi
\expandafter\ifx\csname bibnamefont\endcsname\relax
  \def\bibnamefont#1{#1}\fi
\expandafter\ifx\csname bibfnamefont\endcsname\relax
  \def\bibfnamefont#1{#1}\fi
\expandafter\ifx\csname citenamefont\endcsname\relax
  \def\citenamefont#1{#1}\fi
\expandafter\ifx\csname url\endcsname\relax
  \def\url#1{\texttt{#1}}\fi
\expandafter\ifx\csname urlprefix\endcsname\relax\def\urlprefix{URL }\fi
\providecommand{\bibinfo}[2]{#2}
\providecommand{\eprint}[2][]{\url{#2}}

\bibitem[{\citenamefont{Vanderbilt}(2018)}]{Vanderbilt2018}
\bibinfo{author}{\bibfnamefont{D.}~\bibnamefont{Vanderbilt}},
  \emph{\bibinfo{title}{Berry Phases in Electronic Structure Theory: Electric
  Polarization, Orbital Magnetization and Topological Insulators}}
  (\bibinfo{publisher}{Cambridge University Press}, \bibinfo{year}{2018}).

\bibitem[{\citenamefont{Xiao et~al.}(2010)\citenamefont{Xiao, Chang, and
  Niu}}]{Di_Chang_Niu2010}
\bibinfo{author}{\bibfnamefont{D.}~\bibnamefont{Xiao}},
  \bibinfo{author}{\bibfnamefont{M.-C.} \bibnamefont{Chang}}, \bibnamefont{and}
  \bibinfo{author}{\bibfnamefont{Q.}~\bibnamefont{Niu}}, \bibinfo{journal}{Rev.
  Mod. Phys.} \textbf{\bibinfo{volume}{82}}, \bibinfo{pages}{1959}
  (\bibinfo{year}{2010}),
  \urlprefix\url{https://link.aps.org/doi/10.1103/RevModPhys.82.1959}.

\bibitem[{\citenamefont{Berry}(1984)}]{Berry1984}
\bibinfo{author}{\bibfnamefont{M.~V.} \bibnamefont{Berry}},
  \bibinfo{journal}{Proceedings of the Royal Society of London. A. Mathematical
  and Physical Sciences} \textbf{\bibinfo{volume}{392}}, \bibinfo{pages}{45}
  (\bibinfo{year}{1984}),
  \urlprefix\url{https://royalsocietypublishing.org/doi/abs/10.1098/rspa.1984.0023}.

\bibitem[{\citenamefont{Thouless et~al.}(1982)\citenamefont{Thouless, Kohmoto,
  Nightingale, and den Nijs}}]{TKNN1982}
\bibinfo{author}{\bibfnamefont{D.~J.} \bibnamefont{Thouless}},
  \bibinfo{author}{\bibfnamefont{M.}~\bibnamefont{Kohmoto}},
  \bibinfo{author}{\bibfnamefont{M.~P.} \bibnamefont{Nightingale}},
  \bibnamefont{and} \bibinfo{author}{\bibfnamefont{M.}~\bibnamefont{den Nijs}},
  \bibinfo{journal}{Phys. Rev. Lett.} \textbf{\bibinfo{volume}{49}},
  \bibinfo{pages}{405} (\bibinfo{year}{1982}),
  \urlprefix\url{https://link.aps.org/doi/10.1103/PhysRevLett.49.405}.

\bibitem[{\citenamefont{Vafek and Vishwanath}(2014)}]{Vafek_Vishwanath2014}
\bibinfo{author}{\bibfnamefont{O.}~\bibnamefont{Vafek}} \bibnamefont{and}
  \bibinfo{author}{\bibfnamefont{A.}~\bibnamefont{Vishwanath}},
  \bibinfo{journal}{Annual Review of Condensed Matter Physics}
  \textbf{\bibinfo{volume}{5}}, \bibinfo{pages}{83} (\bibinfo{year}{2014}),
  \urlprefix\url{https://doi.org/10.1146/annurev-conmatphys-031113-133841}.

\bibitem[{\citenamefont{Zhang et~al.}(2006)\citenamefont{Zhang, Jiang, Small,
  Purewal, Tan, Fazlollahi, Chudow, Jaszczak, Stormer, and
  Kim}}]{Zhang_etal2006}
\bibinfo{author}{\bibfnamefont{Y.}~\bibnamefont{Zhang}},
  \bibinfo{author}{\bibfnamefont{Z.}~\bibnamefont{Jiang}},
  \bibinfo{author}{\bibfnamefont{J.~P.} \bibnamefont{Small}},
  \bibinfo{author}{\bibfnamefont{M.~S.} \bibnamefont{Purewal}},
  \bibinfo{author}{\bibfnamefont{Y.-W.} \bibnamefont{Tan}},
  \bibinfo{author}{\bibfnamefont{M.}~\bibnamefont{Fazlollahi}},
  \bibinfo{author}{\bibfnamefont{J.~D.} \bibnamefont{Chudow}},
  \bibinfo{author}{\bibfnamefont{J.~A.} \bibnamefont{Jaszczak}},
  \bibinfo{author}{\bibfnamefont{H.~L.} \bibnamefont{Stormer}},
  \bibnamefont{and} \bibinfo{author}{\bibfnamefont{P.}~\bibnamefont{Kim}},
  \bibinfo{journal}{Phys. Rev. Lett.} \textbf{\bibinfo{volume}{96}},
  \bibinfo{pages}{136806} (\bibinfo{year}{2006}),
  \urlprefix\url{https://link.aps.org/doi/10.1103/PhysRevLett.96.136806}.

\bibitem[{\citenamefont{Novoselov et~al.}(2006)\citenamefont{Novoselov, McCann,
  Morozov, Fal'ko, Katsnelson, Zeitler, Jiang, Schedin, and
  Geim}}]{Novoselov_etal2006}
\bibinfo{author}{\bibfnamefont{K.~S.} \bibnamefont{Novoselov}},
  \bibinfo{author}{\bibfnamefont{E.}~\bibnamefont{McCann}},
  \bibinfo{author}{\bibfnamefont{S.~V.} \bibnamefont{Morozov}},
  \bibinfo{author}{\bibfnamefont{V.~I.} \bibnamefont{Fal'ko}},
  \bibinfo{author}{\bibfnamefont{M.~I.} \bibnamefont{Katsnelson}},
  \bibinfo{author}{\bibfnamefont{U.}~\bibnamefont{Zeitler}},
  \bibinfo{author}{\bibfnamefont{D.}~\bibnamefont{Jiang}},
  \bibinfo{author}{\bibfnamefont{F.}~\bibnamefont{Schedin}}, \bibnamefont{and}
  \bibinfo{author}{\bibfnamefont{A.~K.} \bibnamefont{Geim}},
  \bibinfo{journal}{Nature Physics} \textbf{\bibinfo{volume}{2}},
  \bibinfo{pages}{177} (\bibinfo{year}{2006}),
  \urlprefix\url{https://doi.org/10.1038/nphys245}.

\bibitem[{SU2({\natexlab{a}})}]{SU2_simple_example}
\bibinfo{note}{A simple example is a quantum spin-$\frac{1}{2}$
  $\hat{\mathbf{S}}$ in an external (classical) magnetic field $H = \mathbf{B}
  \cdot \hat{\mathbf{S}}$.}

\bibitem[{gra()}]{graphene_symmetries}
\bibinfo{note}{E.g. in Graphene, space inversion and time reversal get rid of
  one of the parameters \onlinecite{Vafek_Vishwanath2014}.}

\bibitem[{lin()}]{line_degeneracy_berry_phase}
\bibinfo{note}{The wavefunctions in fact do not return to themselves when the
  parameter completes a circuit, unlike the usual case in any (Abelian) Berry
  phase calculation.}

\bibitem[{\citenamefont{Park and Marzari}(2011)}]{Park_Marzari2011}
\bibinfo{author}{\bibfnamefont{C.-H.} \bibnamefont{Park}} \bibnamefont{and}
  \bibinfo{author}{\bibfnamefont{N.}~\bibnamefont{Marzari}},
  \bibinfo{journal}{Phys. Rev. B} \textbf{\bibinfo{volume}{84}},
  \bibinfo{pages}{205440} (\bibinfo{year}{2011}),
  \urlprefix\url{https://link.aps.org/doi/10.1103/PhysRevB.84.205440}.

\bibitem[{\citenamefont{Lin and Liu}(2015)}]{Lin_Liu2015}
\bibinfo{author}{\bibfnamefont{Z.}~\bibnamefont{Lin}} \bibnamefont{and}
  \bibinfo{author}{\bibfnamefont{Z.}~\bibnamefont{Liu}}, \bibinfo{journal}{The
  Journal of Chemical Physics} \textbf{\bibinfo{volume}{143}},
  \bibinfo{pages}{214109} (\bibinfo{year}{2015}),
  \urlprefix\url{https://doi.org/10.1063/1.4936774}.

\bibitem[{\citenamefont{Bradlyn et~al.}(2016)\citenamefont{Bradlyn, Cano, Wang,
  Vergniory, Felser, Cava, and Bernevig}}]{Bradlyn_etal2016}
\bibinfo{author}{\bibfnamefont{B.}~\bibnamefont{Bradlyn}},
  \bibinfo{author}{\bibfnamefont{J.}~\bibnamefont{Cano}},
  \bibinfo{author}{\bibfnamefont{Z.}~\bibnamefont{Wang}},
  \bibinfo{author}{\bibfnamefont{M.~G.} \bibnamefont{Vergniory}},
  \bibinfo{author}{\bibfnamefont{C.}~\bibnamefont{Felser}},
  \bibinfo{author}{\bibfnamefont{R.~J.} \bibnamefont{Cava}}, \bibnamefont{and}
  \bibinfo{author}{\bibfnamefont{B.~A.} \bibnamefont{Bernevig}},
  \bibinfo{journal}{Science} \textbf{\bibinfo{volume}{353}}
  (\bibinfo{year}{2016}), ISSN \bibinfo{issn}{0036-8075},
  \urlprefix\url{https://science.sciencemag.org/content/353/6299/aaf5037}.

\bibitem[{\citenamefont{Green et~al.}(2010)\citenamefont{Green, Santos, and
  Chamon}}]{Green_Santos_Chamon2010}
\bibinfo{author}{\bibfnamefont{D.}~\bibnamefont{Green}},
  \bibinfo{author}{\bibfnamefont{L.}~\bibnamefont{Santos}}, \bibnamefont{and}
  \bibinfo{author}{\bibfnamefont{C.}~\bibnamefont{Chamon}},
  \bibinfo{journal}{Phys. Rev. B} \textbf{\bibinfo{volume}{82}},
  \bibinfo{pages}{075104} (\bibinfo{year}{2010}),
  \urlprefix\url{https://link.aps.org/doi/10.1103/PhysRevB.82.075104}.

\bibitem[{\citenamefont{D\'ora et~al.}(2011)\citenamefont{D\'ora, Kailasvuori,
  and Moessner}}]{Dora_Kailasvuori_Moessner2011}
\bibinfo{author}{\bibfnamefont{B.}~\bibnamefont{D\'ora}},
  \bibinfo{author}{\bibfnamefont{J.}~\bibnamefont{Kailasvuori}},
  \bibnamefont{and} \bibinfo{author}{\bibfnamefont{R.}~\bibnamefont{Moessner}},
  \bibinfo{journal}{Phys. Rev. B} \textbf{\bibinfo{volume}{84}},
  \bibinfo{pages}{195422} (\bibinfo{year}{2011}),
  \urlprefix\url{https://link.aps.org/doi/10.1103/PhysRevB.84.195422}.

\bibitem[{\citenamefont{Lan et~al.}(2011)\citenamefont{Lan, Goldman, Bermudez,
  Lu, and \"Ohberg}}]{Lan_etal2011}
\bibinfo{author}{\bibfnamefont{Z.}~\bibnamefont{Lan}},
  \bibinfo{author}{\bibfnamefont{N.}~\bibnamefont{Goldman}},
  \bibinfo{author}{\bibfnamefont{A.}~\bibnamefont{Bermudez}},
  \bibinfo{author}{\bibfnamefont{W.}~\bibnamefont{Lu}}, \bibnamefont{and}
  \bibinfo{author}{\bibfnamefont{P.}~\bibnamefont{\"Ohberg}},
  \bibinfo{journal}{Phys. Rev. B} \textbf{\bibinfo{volume}{84}},
  \bibinfo{pages}{165115} (\bibinfo{year}{2011}),
  \urlprefix\url{https://link.aps.org/doi/10.1103/PhysRevB.84.165115}.

\bibitem[{\citenamefont{Urban et~al.}(2011)\citenamefont{Urban, Bercioux,
  Wimmer, and H\"ausler}}]{Urban_etal2011}
\bibinfo{author}{\bibfnamefont{D.~F.} \bibnamefont{Urban}},
  \bibinfo{author}{\bibfnamefont{D.}~\bibnamefont{Bercioux}},
  \bibinfo{author}{\bibfnamefont{M.}~\bibnamefont{Wimmer}}, \bibnamefont{and}
  \bibinfo{author}{\bibfnamefont{W.}~\bibnamefont{H\"ausler}},
  \bibinfo{journal}{Phys. Rev. B} \textbf{\bibinfo{volume}{84}},
  \bibinfo{pages}{115136} (\bibinfo{year}{2011}),
  \urlprefix\url{https://link.aps.org/doi/10.1103/PhysRevB.84.115136}.

\bibitem[{\citenamefont{Wang et~al.}(2013)\citenamefont{Wang, Huang, Duan, and
  Liu}}]{Wang_etal2013}
\bibinfo{author}{\bibfnamefont{J.}~\bibnamefont{Wang}},
  \bibinfo{author}{\bibfnamefont{H.}~\bibnamefont{Huang}},
  \bibinfo{author}{\bibfnamefont{W.}~\bibnamefont{Duan}}, \bibnamefont{and}
  \bibinfo{author}{\bibfnamefont{Z.}~\bibnamefont{Liu}}, \bibinfo{journal}{The
  Journal of Chemical Physics} \textbf{\bibinfo{volume}{139}},
  \bibinfo{pages}{184701} (\bibinfo{year}{2013}),
  \urlprefix\url{https://doi.org/10.1063/1.4828861}.

\bibitem[{\citenamefont{Raoux et~al.}(2014)\citenamefont{Raoux, Morigi, Fuchs,
  Pi\'echon, and Montambaux}}]{Raoux_etal2014}
\bibinfo{author}{\bibfnamefont{A.}~\bibnamefont{Raoux}},
  \bibinfo{author}{\bibfnamefont{M.}~\bibnamefont{Morigi}},
  \bibinfo{author}{\bibfnamefont{J.-N.} \bibnamefont{Fuchs}},
  \bibinfo{author}{\bibfnamefont{F.}~\bibnamefont{Pi\'echon}},
  \bibnamefont{and}
  \bibinfo{author}{\bibfnamefont{G.}~\bibnamefont{Montambaux}},
  \bibinfo{journal}{Phys. Rev. Lett.} \textbf{\bibinfo{volume}{112}},
  \bibinfo{pages}{026402} (\bibinfo{year}{2014}),
  \urlprefix\url{https://link.aps.org/doi/10.1103/PhysRevLett.112.026402}.

\bibitem[{\citenamefont{Giovannetti et~al.}(2015)\citenamefont{Giovannetti,
  Capone, van~den Brink, and Ortix}}]{Giovannetti_etal2015}
\bibinfo{author}{\bibfnamefont{G.}~\bibnamefont{Giovannetti}},
  \bibinfo{author}{\bibfnamefont{M.}~\bibnamefont{Capone}},
  \bibinfo{author}{\bibfnamefont{J.}~\bibnamefont{van~den Brink}},
  \bibnamefont{and} \bibinfo{author}{\bibfnamefont{C.}~\bibnamefont{Ortix}},
  \bibinfo{journal}{Phys. Rev. B} \textbf{\bibinfo{volume}{91}},
  \bibinfo{pages}{121417} (\bibinfo{year}{2015}),
  \urlprefix\url{https://link.aps.org/doi/10.1103/PhysRevB.91.121417}.

\bibitem[{\citenamefont{Palumbo and
  Meichanetzidis}(2015)}]{Palumbo_Meichanetzidis2015}
\bibinfo{author}{\bibfnamefont{G.}~\bibnamefont{Palumbo}} \bibnamefont{and}
  \bibinfo{author}{\bibfnamefont{K.}~\bibnamefont{Meichanetzidis}},
  \bibinfo{journal}{Phys. Rev. B} \textbf{\bibinfo{volume}{92}},
  \bibinfo{pages}{235106} (\bibinfo{year}{2015}),
  \urlprefix\url{https://link.aps.org/doi/10.1103/PhysRevB.92.235106}.

\bibitem[{\citenamefont{Xu and Duan}(2017)}]{Xu_Duan2017}
\bibinfo{author}{\bibfnamefont{Y.}~\bibnamefont{Xu}} \bibnamefont{and}
  \bibinfo{author}{\bibfnamefont{L.-M.} \bibnamefont{Duan}},
  \bibinfo{journal}{Phys. Rev. B} \textbf{\bibinfo{volume}{96}},
  \bibinfo{pages}{155301} (\bibinfo{year}{2017}),
  \urlprefix\url{https://link.aps.org/doi/10.1103/PhysRevB.96.155301}.

\bibitem[{\citenamefont{Wang and Yao}(2018)}]{Wang_Yao2018}
\bibinfo{author}{\bibfnamefont{L.}~\bibnamefont{Wang}} \bibnamefont{and}
  \bibinfo{author}{\bibfnamefont{D.-X.} \bibnamefont{Yao}},
  \bibinfo{journal}{Phys. Rev. B} \textbf{\bibinfo{volume}{98}},
  \bibinfo{pages}{161403} (\bibinfo{year}{2018}),
  \urlprefix\url{https://link.aps.org/doi/10.1103/PhysRevB.98.161403}.

\bibitem[{\citenamefont{Lv et~al.}(2017)\citenamefont{Lv, Feng, Xu, Gao, Ma,
  Kong, Richard, Huang, Strocov, Fang et~al.}}]{Lv_etal2017}
\bibinfo{author}{\bibfnamefont{B.~Q.} \bibnamefont{Lv}},
  \bibinfo{author}{\bibfnamefont{Z.-L.} \bibnamefont{Feng}},
  \bibinfo{author}{\bibfnamefont{Q.-N.} \bibnamefont{Xu}},
  \bibinfo{author}{\bibfnamefont{X.}~\bibnamefont{Gao}},
  \bibinfo{author}{\bibfnamefont{J.-Z.} \bibnamefont{Ma}},
  \bibinfo{author}{\bibfnamefont{L.-Y.} \bibnamefont{Kong}},
  \bibinfo{author}{\bibfnamefont{P.}~\bibnamefont{Richard}},
  \bibinfo{author}{\bibfnamefont{Y.-B.} \bibnamefont{Huang}},
  \bibinfo{author}{\bibfnamefont{V.~N.} \bibnamefont{Strocov}},
  \bibinfo{author}{\bibfnamefont{C.}~\bibnamefont{Fang}}, \bibnamefont{et~al.},
  \bibinfo{journal}{Nature} \textbf{\bibinfo{volume}{546}},
  \bibinfo{pages}{627} (\bibinfo{year}{2017}),
  \urlprefix\url{https://doi.org/10.1038/nature22390}.

\bibitem[{\citenamefont{Winkler et~al.}(2016)\citenamefont{Winkler, Wu, Troyer,
  Krogstrup, and Soluyanov}}]{Wrinkler_etal2016}
\bibinfo{author}{\bibfnamefont{G.~W.} \bibnamefont{Winkler}},
  \bibinfo{author}{\bibfnamefont{Q.}~\bibnamefont{Wu}},
  \bibinfo{author}{\bibfnamefont{M.}~\bibnamefont{Troyer}},
  \bibinfo{author}{\bibfnamefont{P.}~\bibnamefont{Krogstrup}},
  \bibnamefont{and} \bibinfo{author}{\bibfnamefont{A.~A.}
  \bibnamefont{Soluyanov}}, \bibinfo{journal}{Phys. Rev. Lett.}
  \textbf{\bibinfo{volume}{117}}, \bibinfo{pages}{076403}
  (\bibinfo{year}{2016}),
  \urlprefix\url{https://link.aps.org/doi/10.1103/PhysRevLett.117.076403}.

\bibitem[{\citenamefont{Weng et~al.}(2016{\natexlab{a}})\citenamefont{Weng,
  Fang, Fang, and Dai}}]{Weng_etal2016a}
\bibinfo{author}{\bibfnamefont{H.}~\bibnamefont{Weng}},
  \bibinfo{author}{\bibfnamefont{C.}~\bibnamefont{Fang}},
  \bibinfo{author}{\bibfnamefont{Z.}~\bibnamefont{Fang}}, \bibnamefont{and}
  \bibinfo{author}{\bibfnamefont{X.}~\bibnamefont{Dai}},
  \bibinfo{journal}{Phys. Rev. B} \textbf{\bibinfo{volume}{93}},
  \bibinfo{pages}{241202} (\bibinfo{year}{2016}{\natexlab{a}}),
  \urlprefix\url{https://link.aps.org/doi/10.1103/PhysRevB.93.241202}.

\bibitem[{\citenamefont{Zhu et~al.}(2016)\citenamefont{Zhu, Winkler, Wu, Li,
  and Soluyanov}}]{Zhu_etal2016}
\bibinfo{author}{\bibfnamefont{Z.}~\bibnamefont{Zhu}},
  \bibinfo{author}{\bibfnamefont{G.~W.} \bibnamefont{Winkler}},
  \bibinfo{author}{\bibfnamefont{Q.}~\bibnamefont{Wu}},
  \bibinfo{author}{\bibfnamefont{J.}~\bibnamefont{Li}}, \bibnamefont{and}
  \bibinfo{author}{\bibfnamefont{A.~A.} \bibnamefont{Soluyanov}},
  \bibinfo{journal}{Phys. Rev. X} \textbf{\bibinfo{volume}{6}},
  \bibinfo{pages}{031003} (\bibinfo{year}{2016}),
  \urlprefix\url{https://link.aps.org/doi/10.1103/PhysRevX.6.031003}.

\bibitem[{\citenamefont{Weng et~al.}(2016{\natexlab{b}})\citenamefont{Weng,
  Fang, Fang, and Dai}}]{Weng_etal2016b}
\bibinfo{author}{\bibfnamefont{H.}~\bibnamefont{Weng}},
  \bibinfo{author}{\bibfnamefont{C.}~\bibnamefont{Fang}},
  \bibinfo{author}{\bibfnamefont{Z.}~\bibnamefont{Fang}}, \bibnamefont{and}
  \bibinfo{author}{\bibfnamefont{X.}~\bibnamefont{Dai}},
  \bibinfo{journal}{Phys. Rev. B} \textbf{\bibinfo{volume}{94}},
  \bibinfo{pages}{165201} (\bibinfo{year}{2016}{\natexlab{b}}),
  \urlprefix\url{https://link.aps.org/doi/10.1103/PhysRevB.94.165201}.

\bibitem[{\citenamefont{Chang et~al.}(2017{\natexlab{a}})\citenamefont{Chang,
  Xu, Huang, Sanchez, Hsu, Bian, Yu, Belopolski, Alidoust, Zheng
  et~al.}}]{Chang_etal2017a}
\bibinfo{author}{\bibfnamefont{G.}~\bibnamefont{Chang}},
  \bibinfo{author}{\bibfnamefont{S.-Y.} \bibnamefont{Xu}},
  \bibinfo{author}{\bibfnamefont{S.-M.} \bibnamefont{Huang}},
  \bibinfo{author}{\bibfnamefont{D.~S.} \bibnamefont{Sanchez}},
  \bibinfo{author}{\bibfnamefont{C.-H.} \bibnamefont{Hsu}},
  \bibinfo{author}{\bibfnamefont{G.}~\bibnamefont{Bian}},
  \bibinfo{author}{\bibfnamefont{Z.-M.} \bibnamefont{Yu}},
  \bibinfo{author}{\bibfnamefont{I.}~\bibnamefont{Belopolski}},
  \bibinfo{author}{\bibfnamefont{N.}~\bibnamefont{Alidoust}},
  \bibinfo{author}{\bibfnamefont{H.}~\bibnamefont{Zheng}},
  \bibnamefont{et~al.}, \bibinfo{journal}{Scientific Reports}
  \textbf{\bibinfo{volume}{7}}, \bibinfo{pages}{1688}
  (\bibinfo{year}{2017}{\natexlab{a}}), ISSN \bibinfo{issn}{2045-2322},
  \urlprefix\url{https://doi.org/10.1038/s41598-017-01523-8}.

\bibitem[{\citenamefont{Fulga and Stern}(2017)}]{Fulga_Stern2017}
\bibinfo{author}{\bibfnamefont{I.~C.} \bibnamefont{Fulga}} \bibnamefont{and}
  \bibinfo{author}{\bibfnamefont{A.}~\bibnamefont{Stern}},
  \bibinfo{journal}{Phys. Rev. B} \textbf{\bibinfo{volume}{95}},
  \bibinfo{pages}{241116} (\bibinfo{year}{2017}),
  \urlprefix\url{https://link.aps.org/doi/10.1103/PhysRevB.95.241116}.

\bibitem[{\citenamefont{Chang et~al.}(2017{\natexlab{b}})\citenamefont{Chang,
  Xu, Wieder, Sanchez, Huang, Belopolski, Chang, Zhang, Bansil, Lin
  et~al.}}]{Chang_etal2017b}
\bibinfo{author}{\bibfnamefont{G.}~\bibnamefont{Chang}},
  \bibinfo{author}{\bibfnamefont{S.-Y.} \bibnamefont{Xu}},
  \bibinfo{author}{\bibfnamefont{B.~J.} \bibnamefont{Wieder}},
  \bibinfo{author}{\bibfnamefont{D.~S.} \bibnamefont{Sanchez}},
  \bibinfo{author}{\bibfnamefont{S.-M.} \bibnamefont{Huang}},
  \bibinfo{author}{\bibfnamefont{I.}~\bibnamefont{Belopolski}},
  \bibinfo{author}{\bibfnamefont{T.-R.} \bibnamefont{Chang}},
  \bibinfo{author}{\bibfnamefont{S.}~\bibnamefont{Zhang}},
  \bibinfo{author}{\bibfnamefont{A.}~\bibnamefont{Bansil}},
  \bibinfo{author}{\bibfnamefont{H.}~\bibnamefont{Lin}}, \bibnamefont{et~al.},
  \bibinfo{journal}{Phys. Rev. Lett.} \textbf{\bibinfo{volume}{119}},
  \bibinfo{pages}{206401} (\bibinfo{year}{2017}{\natexlab{b}}),
  \urlprefix\url{https://link.aps.org/doi/10.1103/PhysRevLett.119.206401}.

\bibitem[{\citenamefont{Zhong et~al.}(2017)\citenamefont{Zhong, Chen, Yu, Xie,
  Wang, Yang, and Zhang}}]{Zhong_etal2017}
\bibinfo{author}{\bibfnamefont{C.}~\bibnamefont{Zhong}},
  \bibinfo{author}{\bibfnamefont{Y.}~\bibnamefont{Chen}},
  \bibinfo{author}{\bibfnamefont{Z.-M.} \bibnamefont{Yu}},
  \bibinfo{author}{\bibfnamefont{Y.}~\bibnamefont{Xie}},
  \bibinfo{author}{\bibfnamefont{H.}~\bibnamefont{Wang}},
  \bibinfo{author}{\bibfnamefont{S.~A.} \bibnamefont{Yang}}, \bibnamefont{and}
  \bibinfo{author}{\bibfnamefont{S.}~\bibnamefont{Zhang}},
  \bibinfo{journal}{Nature Communications} \textbf{\bibinfo{volume}{8}},
  \bibinfo{pages}{15641 EP } (\bibinfo{year}{2017}), \bibinfo{note}{article},
  \urlprefix\url{https://doi.org/10.1038/ncomms15641}.

\bibitem[{\citenamefont{Yu et~al.}(2017)\citenamefont{Yu, Yan, and
  Liu}}]{Yu_Yan_Liu2017}
\bibinfo{author}{\bibfnamefont{J.}~\bibnamefont{Yu}},
  \bibinfo{author}{\bibfnamefont{B.}~\bibnamefont{Yan}}, \bibnamefont{and}
  \bibinfo{author}{\bibfnamefont{C.-X.} \bibnamefont{Liu}},
  \bibinfo{journal}{Phys. Rev. B} \textbf{\bibinfo{volume}{95}},
  \bibinfo{pages}{235158} (\bibinfo{year}{2017}),
  \urlprefix\url{https://link.aps.org/doi/10.1103/PhysRevB.95.235158}.

\bibitem[{\citenamefont{Zhang et~al.}(2017)\citenamefont{Zhang, Yu, Sheng,
  Yang, and Yang}}]{Zhang_etal2017}
\bibinfo{author}{\bibfnamefont{X.}~\bibnamefont{Zhang}},
  \bibinfo{author}{\bibfnamefont{Z.-M.} \bibnamefont{Yu}},
  \bibinfo{author}{\bibfnamefont{X.-L.} \bibnamefont{Sheng}},
  \bibinfo{author}{\bibfnamefont{H.~Y.} \bibnamefont{Yang}}, \bibnamefont{and}
  \bibinfo{author}{\bibfnamefont{S.~A.} \bibnamefont{Yang}},
  \bibinfo{journal}{Phys. Rev. B} \textbf{\bibinfo{volume}{95}},
  \bibinfo{pages}{235116} (\bibinfo{year}{2017}),
  \urlprefix\url{https://link.aps.org/doi/10.1103/PhysRevB.95.235116}.

\bibitem[{\citenamefont{Yang et~al.}(2017)\citenamefont{Yang, Yu, Parkin,
  Felser, Liu, and Yan}}]{Yang_etal2017}
\bibinfo{author}{\bibfnamefont{H.}~\bibnamefont{Yang}},
  \bibinfo{author}{\bibfnamefont{J.}~\bibnamefont{Yu}},
  \bibinfo{author}{\bibfnamefont{S.~S.~P.} \bibnamefont{Parkin}},
  \bibinfo{author}{\bibfnamefont{C.}~\bibnamefont{Felser}},
  \bibinfo{author}{\bibfnamefont{C.-X.} \bibnamefont{Liu}}, \bibnamefont{and}
  \bibinfo{author}{\bibfnamefont{B.}~\bibnamefont{Yan}},
  \bibinfo{journal}{Phys. Rev. Lett.} \textbf{\bibinfo{volume}{119}},
  \bibinfo{pages}{136401} (\bibinfo{year}{2017}),
  \urlprefix\url{https://link.aps.org/doi/10.1103/PhysRevLett.119.136401}.

\bibitem[{\citenamefont{Wallace}(1947)}]{Wallace1947}
\bibinfo{author}{\bibfnamefont{P.~R.} \bibnamefont{Wallace}},
  \bibinfo{journal}{Physical Review} \textbf{\bibinfo{volume}{71}},
  \bibinfo{pages}{622} (\bibinfo{year}{1947}),
  \urlprefix\url{https://doi.org/10.1103/physrev.71.622}.

\bibitem[{SU2({\natexlab{b}})}]{SU2_spin1}
\bibinfo{note}{Essentially, $H^{3B}_K$ is equivalent to $\mathbf{B} \cdot
  \hat{\mathbf{S}}$ for a quantum spin-1 in an external (classical) magnetic
  field that is confined to a plane.}

\bibitem[{\citenamefont{Halzen and Martin}(1984)}]{Halzen_Martin1984}
\bibinfo{author}{\bibfnamefont{F.}~\bibnamefont{Halzen}} \bibnamefont{and}
  \bibinfo{author}{\bibfnamefont{A.~D.} \bibnamefont{Martin}},
  \emph{\bibinfo{title}{{Quarks and Leptons: An introductory course in modern
  particle physics}}} (\bibinfo{publisher}{New York, USA: Wiley},
  \bibinfo{year}{1984}), ISBN \bibinfo{isbn}{0471887412, 9780471887416}.

\bibitem[{cas()}]{case3a_footnote}
\bibinfo{note}{In terms of the triplet of indices introduced in Sec.
  \ref{sec:3band}, this two-fold degeneracy gets classified as 1
  ($\theta'_{\mathbf{p}}$), 1 ($H$) and 0(t), 2(m), 2(b), where the prime on
  $\theta'_{\mathbf{p}}$ is to point out that the circuit is made near this
  two-fold degeneracy which is not at the origin.}

\bibitem[{\citenamefont{Mikitik and Sharlai}(2008)}]{Mikitik_Sharlai2008}
\bibinfo{author}{\bibfnamefont{G.~P.} \bibnamefont{Mikitik}} \bibnamefont{and}
  \bibinfo{author}{\bibfnamefont{Y.~V.} \bibnamefont{Sharlai}},
  \bibinfo{journal}{Phys. Rev. B} \textbf{\bibinfo{volume}{77}},
  \bibinfo{pages}{113407} (\bibinfo{year}{2008}),
  \urlprefix\url{https://link.aps.org/doi/10.1103/PhysRevB.77.113407}.

\bibitem[{\citenamefont{Fuchs et~al.}(2010)\citenamefont{Fuchs, Pi{\'{e}}chon,
  Goerbig, and Montambaux}}]{Fuchs_etal2010}
\bibinfo{author}{\bibfnamefont{J.~N.} \bibnamefont{Fuchs}},
  \bibinfo{author}{\bibfnamefont{F.}~\bibnamefont{Pi{\'{e}}chon}},
  \bibinfo{author}{\bibfnamefont{M.~O.} \bibnamefont{Goerbig}},
  \bibnamefont{and}
  \bibinfo{author}{\bibfnamefont{G.}~\bibnamefont{Montambaux}},
  \bibinfo{journal}{The European Physical Journal B}
  \textbf{\bibinfo{volume}{77}}, \bibinfo{pages}{351} (\bibinfo{year}{2010}),
  \urlprefix\url{https://doi.org/10.1140/epjb/e2010-00259-2}.

\bibitem[{\citenamefont{Ozerin and Falkovsky}(2012)}]{Ozerin_Falkovsky2012}
\bibinfo{author}{\bibfnamefont{A.~Y.} \bibnamefont{Ozerin}} \bibnamefont{and}
  \bibinfo{author}{\bibfnamefont{L.~A.} \bibnamefont{Falkovsky}},
  \bibinfo{journal}{Physical Review B} \textbf{\bibinfo{volume}{85}}
  (\bibinfo{year}{2012}),
  \urlprefix\url{https://doi.org/10.1103/physrevb.85.205143}.

\bibitem[{\citenamefont{Ryu et~al.}(2010)\citenamefont{Ryu, Schnyder, Furusaki,
  and Ludwig}}]{Ryu_2010}
\bibinfo{author}{\bibfnamefont{S.}~\bibnamefont{Ryu}},
  \bibinfo{author}{\bibfnamefont{A.~P.} \bibnamefont{Schnyder}},
  \bibinfo{author}{\bibfnamefont{A.}~\bibnamefont{Furusaki}}, \bibnamefont{and}
  \bibinfo{author}{\bibfnamefont{A.~W.~W.} \bibnamefont{Ludwig}},
  \bibinfo{journal}{New Journal of Physics} \textbf{\bibinfo{volume}{12}},
  \bibinfo{pages}{065010} (\bibinfo{year}{2010}),
  \urlprefix\url{https://doi.org/10.1088%2F1367-2630%2F12%2F6%2F065010}.

\bibitem[{\citenamefont{Barman et~al.}(2019)\citenamefont{Barman, Mondal,
  Pathak, and Alam}}]{Barman_etal2019}
\bibinfo{author}{\bibfnamefont{C.~K.} \bibnamefont{Barman}},
  \bibinfo{author}{\bibfnamefont{C.}~\bibnamefont{Mondal}},
  \bibinfo{author}{\bibfnamefont{B.}~\bibnamefont{Pathak}}, \bibnamefont{and}
  \bibinfo{author}{\bibfnamefont{A.}~\bibnamefont{Alam}},
  \bibinfo{journal}{Phys. Rev. B} \textbf{\bibinfo{volume}{99}},
  \bibinfo{pages}{045144} (\bibinfo{year}{2019}),
  \urlprefix\url{https://link.aps.org/doi/10.1103/PhysRevB.99.045144}.

\bibitem[{\citenamefont{Rhim and Park}(2012)}]{Rhim_Park2012}
\bibinfo{author}{\bibfnamefont{J.-W.} \bibnamefont{Rhim}} \bibnamefont{and}
  \bibinfo{author}{\bibfnamefont{K.}~\bibnamefont{Park}},
  \bibinfo{journal}{Phys. Rev. B} \textbf{\bibinfo{volume}{86}},
  \bibinfo{pages}{235411} (\bibinfo{year}{2012}),
  \urlprefix\url{https://link.aps.org/doi/10.1103/PhysRevB.86.235411}.

\bibitem[{\citenamefont{Illes and Nicol}(2016)}]{Illes_Nicol2016}
\bibinfo{author}{\bibfnamefont{E.}~\bibnamefont{Illes}} \bibnamefont{and}
  \bibinfo{author}{\bibfnamefont{E.~J.} \bibnamefont{Nicol}},
  \bibinfo{journal}{Phys. Rev. B} \textbf{\bibinfo{volume}{94}},
  \bibinfo{pages}{125435} (\bibinfo{year}{2016}),
  \urlprefix\url{https://link.aps.org/doi/10.1103/PhysRevB.94.125435}.

\bibitem[{\citenamefont{Kov\'acs et~al.}(2017)\citenamefont{Kov\'acs, D\'avid,
  D\'ora, and Cserti}}]{Kovacs_etal2017}
\bibinfo{author}{\bibfnamefont{A.~D.} \bibnamefont{Kov\'acs}},
  \bibinfo{author}{\bibfnamefont{G.}~\bibnamefont{D\'avid}},
  \bibinfo{author}{\bibfnamefont{B.}~\bibnamefont{D\'ora}}, \bibnamefont{and}
  \bibinfo{author}{\bibfnamefont{J.}~\bibnamefont{Cserti}},
  \bibinfo{journal}{Phys. Rev. B} \textbf{\bibinfo{volume}{95}},
  \bibinfo{pages}{035414} (\bibinfo{year}{2017}),
  \urlprefix\url{https://link.aps.org/doi/10.1103/PhysRevB.95.035414}.

\bibitem[{\citenamefont{Islam and Dutta}(2017)}]{Islam_Dutta2017}
\bibinfo{author}{\bibfnamefont{S.~F.} \bibnamefont{Islam}} \bibnamefont{and}
  \bibinfo{author}{\bibfnamefont{P.}~\bibnamefont{Dutta}},
  \bibinfo{journal}{Phys. Rev. B} \textbf{\bibinfo{volume}{96}},
  \bibinfo{pages}{045418} (\bibinfo{year}{2017}),
  \urlprefix\url{https://link.aps.org/doi/10.1103/PhysRevB.96.045418}.

\bibitem[{\citenamefont{Gorbar et~al.}(2019)\citenamefont{Gorbar, Gusynin, and
  Oriekhov}}]{Gorbar_Gusynin_Oriekhov2019}
\bibinfo{author}{\bibfnamefont{E.~V.} \bibnamefont{Gorbar}},
  \bibinfo{author}{\bibfnamefont{V.~P.} \bibnamefont{Gusynin}},
  \bibnamefont{and} \bibinfo{author}{\bibfnamefont{D.~O.}
  \bibnamefont{Oriekhov}}, \bibinfo{journal}{Phys. Rev. B}
  \textbf{\bibinfo{volume}{99}}, \bibinfo{pages}{155124}
  (\bibinfo{year}{2019}),
  \urlprefix\url{https://link.aps.org/doi/10.1103/PhysRevB.99.155124}.

\bibitem[{\citenamefont{Soukoulis}(2001)}]{Costas2001}
\bibinfo{editor}{\bibfnamefont{C.~M.} \bibnamefont{Soukoulis}}, ed.,
  \emph{\bibinfo{title}{Photonic Crystals and Light Localization in the 21st
  Century}} (\bibinfo{publisher}{Springer Netherlands}, \bibinfo{year}{2001}),
  \urlprefix\url{https://doi.org/10.1007/978-94-010-0738-2}.

\bibitem[{\citenamefont{Ray et~al.}(2017)\citenamefont{Ray, Ghatak, and
  Das}}]{Ray_Ghatak_Das2017}
\bibinfo{author}{\bibfnamefont{S.}~\bibnamefont{Ray}},
  \bibinfo{author}{\bibfnamefont{A.}~\bibnamefont{Ghatak}}, \bibnamefont{and}
  \bibinfo{author}{\bibfnamefont{T.}~\bibnamefont{Das}},
  \bibinfo{journal}{Phys. Rev. B} \textbf{\bibinfo{volume}{95}},
  \bibinfo{pages}{165425} (\bibinfo{year}{2017}),
  \urlprefix\url{https://link.aps.org/doi/10.1103/PhysRevB.95.165425}.

\bibitem[{\citenamefont{Lewenstein et~al.}(2007)\citenamefont{Lewenstein,
  Sanpera, Ahufinger, Damski, Sen(De), and Sen}}]{Lewenstein_etal2007}
\bibinfo{author}{\bibfnamefont{M.}~\bibnamefont{Lewenstein}},
  \bibinfo{author}{\bibfnamefont{A.}~\bibnamefont{Sanpera}},
  \bibinfo{author}{\bibfnamefont{V.}~\bibnamefont{Ahufinger}},
  \bibinfo{author}{\bibfnamefont{B.}~\bibnamefont{Damski}},
  \bibinfo{author}{\bibfnamefont{A.}~\bibnamefont{Sen(De)}}, \bibnamefont{and}
  \bibinfo{author}{\bibfnamefont{U.}~\bibnamefont{Sen}},
  \bibinfo{journal}{Advances in Physics} \textbf{\bibinfo{volume}{56}},
  \bibinfo{pages}{243} (\bibinfo{year}{2007}),
  \urlprefix\url{https://doi.org/10.1080/00018730701223200}.

\bibitem[{\citenamefont{Tarruell et~al.}(2012)\citenamefont{Tarruell, Greif,
  Uehlinger, Jotzu, and Esslinger}}]{Tarruell_etal2012}
\bibinfo{author}{\bibfnamefont{L.}~\bibnamefont{Tarruell}},
  \bibinfo{author}{\bibfnamefont{D.}~\bibnamefont{Greif}},
  \bibinfo{author}{\bibfnamefont{T.}~\bibnamefont{Uehlinger}},
  \bibinfo{author}{\bibfnamefont{G.}~\bibnamefont{Jotzu}}, \bibnamefont{and}
  \bibinfo{author}{\bibfnamefont{T.}~\bibnamefont{Esslinger}},
  \bibinfo{journal}{Nature} \textbf{\bibinfo{volume}{483}},
  \bibinfo{pages}{302} (\bibinfo{year}{2012}),
  \urlprefix\url{https://doi.org/10.1038/nature10871}.

\bibitem[{\citenamefont{Hu et~al.}(2018)\citenamefont{Hu, Hou, Zhang, and
  Zhang}}]{Hu_etal2018}
\bibinfo{author}{\bibfnamefont{H.}~\bibnamefont{Hu}},
  \bibinfo{author}{\bibfnamefont{J.}~\bibnamefont{Hou}},
  \bibinfo{author}{\bibfnamefont{F.}~\bibnamefont{Zhang}}, \bibnamefont{and}
  \bibinfo{author}{\bibfnamefont{C.}~\bibnamefont{Zhang}},
  \bibinfo{journal}{Phys. Rev. Lett.} \textbf{\bibinfo{volume}{120}},
  \bibinfo{pages}{240401} (\bibinfo{year}{2018}),
  \urlprefix\url{https://link.aps.org/doi/10.1103/PhysRevLett.120.240401}.

\bibitem[{\citenamefont{Hu and Zhang}(2018)}]{Hu_Zhang2018}
\bibinfo{author}{\bibfnamefont{H.}~\bibnamefont{Hu}} \bibnamefont{and}
  \bibinfo{author}{\bibfnamefont{C.}~\bibnamefont{Zhang}},
  \bibinfo{journal}{Phys. Rev. A} \textbf{\bibinfo{volume}{98}},
  \bibinfo{pages}{013627} (\bibinfo{year}{2018}),
  \urlprefix\url{https://link.aps.org/doi/10.1103/PhysRevA.98.013627}.

\bibitem[{\citenamefont{Tadjine et~al.}(2016)\citenamefont{Tadjine, Allan, and
  Delerue}}]{Tadjine_Allan_Delerue2016}
\bibinfo{author}{\bibfnamefont{A.}~\bibnamefont{Tadjine}},
  \bibinfo{author}{\bibfnamefont{G.}~\bibnamefont{Allan}}, \bibnamefont{and}
  \bibinfo{author}{\bibfnamefont{C.}~\bibnamefont{Delerue}},
  \bibinfo{journal}{Phys. Rev. B} \textbf{\bibinfo{volume}{94}},
  \bibinfo{pages}{075441} (\bibinfo{year}{2016}),
  \urlprefix\url{https://link.aps.org/doi/10.1103/PhysRevB.94.075441}.

\bibitem[{\citenamefont{Slot et~al.}(2019)\citenamefont{Slot, Kempkes, Knol,
  van Weerdenburg, van~den Broeke, Wegner, Vanmaekelbergh, Khajetoorians,
  Morais~Smith, and Swart}}]{Slot_etal2019}
\bibinfo{author}{\bibfnamefont{M.~R.} \bibnamefont{Slot}},
  \bibinfo{author}{\bibfnamefont{S.~N.} \bibnamefont{Kempkes}},
  \bibinfo{author}{\bibfnamefont{E.~J.} \bibnamefont{Knol}},
  \bibinfo{author}{\bibfnamefont{W.~M.~J.} \bibnamefont{van Weerdenburg}},
  \bibinfo{author}{\bibfnamefont{J.~J.} \bibnamefont{van~den Broeke}},
  \bibinfo{author}{\bibfnamefont{D.}~\bibnamefont{Wegner}},
  \bibinfo{author}{\bibfnamefont{D.}~\bibnamefont{Vanmaekelbergh}},
  \bibinfo{author}{\bibfnamefont{A.~A.} \bibnamefont{Khajetoorians}},
  \bibinfo{author}{\bibfnamefont{C.}~\bibnamefont{Morais~Smith}},
  \bibnamefont{and} \bibinfo{author}{\bibfnamefont{I.}~\bibnamefont{Swart}},
  \bibinfo{journal}{Phys. Rev. X} \textbf{\bibinfo{volume}{9}},
  \bibinfo{pages}{011009} (\bibinfo{year}{2019}),
  \urlprefix\url{https://link.aps.org/doi/10.1103/PhysRevX.9.011009}.

\end{thebibliography}
\end{document}